\documentclass[usenatbib]{mn2e}
\usepackage{amsmath,fleqn}

\usepackage{epsf}

\newcommand\be{{\bmath e}}

\newcommand\bg{{\bmath g}}

\newcommand\bh{{\bmath h}}
\newcommand\bl{{\bmath l}}
\newcommand\bm{{\bmath m}}
\newcommand\bn{{\bmath n}}
\newcommand\bS{{\bmath S}}

\newcommand\bu{{\bmath u}}
\newcommand\bv{{\bmath v}}
\newcommand\bx{{\bmath x}}
\newcommand\bOmega{{\bmath\Omega}}
\newcommand\calF{\mathcal{F}}
\newcommand\calI{\mathcal{I}}
\newcommand\calM{\mathcal{M}}
\newcommand\bcalG{{\bmath{\mathcal{G}}}}
\newcommand\bcalT{{\bmath{\mathcal{T}}}}
\newcommand\real{\mathrm{Re}}

\newcommand\half{{\textstyle\frac{1}{2}}}
\newcommand\third{{\textstyle\frac{1}{3}}}

\newcommand\rmb{\mathrm{b}}

\newcommand\rmd{\mathrm{d}}
\newcommand\rme{\mathrm{e}}

\newcommand\rmh{\mathrm{h}}
\newcommand\rmi{\mathrm{i}}

\newcommand\rmp{\mathrm{p}}
\newcommand\rms{\mathrm{s}}

\newcommand\rmD{\mathrm{D}}

\newcommand\f{\frac}
\newcommand\p{\partial}
\newcommand\cst{\mathrm{constant}}

\title[Local and global dynamics of warped discs]
{Local and global dynamics of warped astrophysical discs}

\author[Gordon I.\ Ogilvie and Henrik N.\ Latter]
{Gordon I.\ Ogilvie and Henrik N.\ Latter\\
Department of Applied Mathematics and Theoretical Physics,
University of Cambridge, Centre for Mathematical Sciences,\\
Wilberforce Road, Cambridge CB3 0WA}

\begin{document}

\maketitle

\label{firstpage}
 
\begin{abstract}
  Astrophysical discs are warped whenever a misalignment is present in
  the system, or when a flat disc is made unstable by external forces.
  The evolution of the shape and mass distribution of a warped disc is
  driven not only by external influences but also by an internal
  torque, which transports angular momentum through the disc.  This
  torque depends on internal flows driven by the oscillating pressure
  gradient associated with the warp, and on physical processes
  operating on smaller scales, which may include instability and
  turbulence.  We introduce a local model for the detailed study of
  warped discs.  Starting from the shearing sheet of Goldreich \&
  Lynden-Bell, we impose the oscillating geometry of the orbital plane
  by means of a coordinate transformation.  This \textit{warped
    shearing sheet} (or \textit{box}) is suitable for analytical and
  computational treatments of fluid dynamics, magnetohydrodynamics,
  etc., and it can be used to compute the internal torque that drives
  the large-scale evolution of the disc.  The simplest hydrodynamic
  states in the local model are horizontally uniform laminar flows
  that oscillate at the orbital frequency.  These correspond to the
  nonlinear solutions for warped discs found in previous work by
  Ogilvie, and we present an alternative derivation and generalization
  of that theory.  In a companion paper we show that these laminar
  flows are often linearly unstable, especially if the disc is nearly
  Keplerian and of low viscosity.  The local model can be used in
  future work to determine the nonlinear outcome of the hydrodynamic
  instability of warped discs, and its interaction with others such as
  the magnetorotational instability.
\end{abstract}


\section{Introduction}

\subsection{Astrophysical motivation}

Warped discs, in which the orbital plane varies with radius, have many
applications in astrophysics.  They occur whenever a misalignment is
present in the system, as in the classic scenario of a black hole
whose spin axis does not coincide with the orbital axis of gas that is
supplied through an accretion disc \citep{1975ApJ...195L..65B}.
Variants of this problem occur when the central object is a magnetized
star or a close binary, interacting with the disc through magnetic or
gravitational torques.  A disc may also be warped by a companion
object on an inclined orbit \citep{1995MNRAS.274..987P}; this
situation is found in sufficiently wide young binary stars and can
occur in protoplanetary systems if a mutual inclination of the planet
and disc is excited by many-body interactions.  Even in systems that
are initially coplanar, warps may arise spontaneously through the
growth of instabilities, such as those involving tidal forces
\citep{1992ApJ...398..525L}, winds \citep{1994A&A...289..149S},
radiation forces \citep{1996MNRAS.281..357P} or magnetic fields
\citep{1999ApJ...524.1030L}.

Early studies of warped discs were motivated not only by theoretical
problems such as misaligned accretion on to a spinning black hole
\citep{1975ApJ...195L..65B}, but also by observational discoveries
such as the low-mass X-ray binary Her X-1 (HZ Her), where the
existence of a precessing disc tilted out of the binary plane was
deduced from light curves \citep{1973NPhS..246...87K}.  Much later,
observations of water masers in the galaxy NGC~4258 (M~106) revealed a
warped disc around the central black hole \citep{1995Natur.373..127M}.
There are by now multiple examples of X-ray binaries
\citep[e.g.][]{2012MNRAS.420.1575K} and active galactic nuclei
\citep[e.g.][]{2005ASPC..340..203G} that may have similar properties
to these systems.  Recent interest in warped discs has focused mainly
on applications to accreting black holes and to protoplanetary
systems.  \citet{2012MNRAS.421.1201N}, \citet{2012MNRAS.422.2547N} and
\citet{2012ApJ...757L..24N} have argued that, in significantly
misaligned accretion on to a spinning black hole, the disc breaks into
rings that can precess independently, and the accretion rate is
greatly enhanced.  \citet{2011MNRAS.412.2799F} have calculated the
warping of a protoplanetary disc that is tilted with respect to the
spin axis of a magnetized central star, and have investigated the
consequences of this dynamics for the spin--orbit misalignment of
extrasolar planetary systems.

\subsection{Theoretical and computational background}

Fundamental theoretical studies of warped discs have mostly aimed to
derive equations that govern the evolution of the shape of the disc
and, in some cases, its mass distribution.  There is also an extensive
literature that applies the theory of warped discs, often in a
simplified form, to astrophysical systems.

Early versions of evolutionary equations for the shape of a warped
disc
\citep{1975ApJ...195L..65B,1977ApJ...214..550P,1978ApJ...226..253P,1981ApJ...247..677H}
differed slightly from each other but all suggested that the warp
would diffuse on a viscous timescale.  They all turned out to be
incorrect, mainly because the internal flows driven by the oscillating
pressure gradient in a warped disc had not been considered.
\citet{1983MNRAS.202.1181P} provided the first consistent linear
theory for viscous Keplerian discs \citep[summarized
by][]{1985MNRAS.213..435K}, and found that the warp diffuses on a
timescale that is shorter than the viscous timescale by a factor of
order $\alpha^2$, where $\alpha\ll1$ is the Shakura--Sunyaev viscosity
parameter.  (In this context, viscosity is usually taken to represent
the effects of unresolved physical processes such as small-scale
turbulence.)  Subsequently, \citet{1995ApJ...438..841P} showed that a
transition from diffusive to wavelike propagation occurs in a
Keplerian disc when $\alpha$ is less than the angular semithickness
$H/r$ of the disc.

\citet{1999MNRAS.304..557O} derived a fully nonlinear theory for the
diffusive regime in Keplerian discs and also for bending waves in
non-Keplerian discs.  While the resulting evolutionary equations are
similar in form to those suggested by \citet{1983MNRAS.202.1181P} and
\citet{1992MNRAS.258..811P}, the analysis also provides a means to
calculate the torque coefficients in those equations as functions of
the amplitude of the warp and other relevant parameters.  The thermal
physics of warped discs in the nonlinear regime was taken into account
by \citet{2000MNRAS.317..607O}.  More recently,
\citet{2006MNRAS.365..977O} showed how the wavelike regime for
Keplerian discs is modified by weak nonlinearity and dispersion:
solitary bending waves would be possible if the adiabatic exponent
$\gamma$ were to exceed~$3$, but otherwise the dominant weakly
nonlinear effect is to enhance the linear dispersion of a bending
wave.

Global numerical simulations of warped discs are very demanding
because of the ranges of length-scales and time-scales that are
involved in a thin disc.  While there have been some grid-based
simulations of warped discs, the majority of studies have used
smoothed particle hydrodynamics (SPH), which is well suited to the
complicated geometry, although less good for resolving small scales.
The evolution of simple warps in SPH simulations was measured and
compared with theoretical expectations by \citet{1999MNRAS.309..929N},
\citet{2007MNRAS.381.1287L} and \citet{2010MNRAS.405.1212L}.  The last
paper, in particular, confirms some aspects of the nonlinear theory of
\citet{1999MNRAS.304..557O}.  Previously, \citet{1996MNRAS.282..597L},
\citet{1997MNRAS.285..288L} and \citet{1997MNRAS.290..490L} had
studied the interaction of discs with binary companions on inclined
orbits; the recent work of \citet{2013MNRAS.431.1320X} involves
planets on inclined orbits.  \citet{2000MNRAS.315..570N} included a
post-Newtonian force within SPH to simulate tilted discs around
spinning black holes.  The tilting and warping of discs in binary
stars by magnetic and radiation forces has been simulated by
\citet{2002MNRAS.335..247M} and
\citet{2006MNRAS.366.1399F,2010MNRAS.401.1275F}.  Simulations using
grid-based methods have been applied to tilted discs around spinning
black holes
\citep{2005ApJ...623..347F,2007ApJ...668..417F,2009ApJ...691..482F},
where they reveal a complicated behaviour of the accretion streams
close to the event horizon.  \citet{2010A&A...511A..77F} used rotating
grids to compute precessing circumstellar discs in binary stars,
obtaining general agreement with theoretical expectations, but finding
that in extreme cases the disc tends to break into independently
precessing rings.  This type of behaviour has also been emphasized in
the previously mentioned work by Nixon
\citep[e.g.][]{2012ApJ...757L..24N}, which uses SPH.

\subsection{Plan of this paper}

The main purpose of this paper is to introduce a local model for the
detailed study of warped discs, including instability and turbulence.
We first discuss the large-scale geometry of a warped disc and show
how the evolution of its shape and mass distribution is driven by an
internal torque.  We then use a circular reference orbit to construct
a standard local model, equivalent to the shearing sheet
\citep{1965MNRAS.130..125G} and including vertical gravity.  To take
into account the oscillating local geometry of the orbital plane we
then introduce a transformation to warped shearing coordinates and
formulate the hydrodynamic equations in this new system.  A single
dimensionless parameter defines the local properties of the warp in
this model, and we show how to compute the internal torque that
governs the large-scale evolution of the disc.  The simplest
hydrodynamic states in the warped shearing box are horizontally
uniform laminar flows that oscillate at the orbital frequency.  We
explore the properties of these laminar flows and the related torques,
which correspond to the nonlinear solutions for warped discs found by
\citet{1999MNRAS.304..557O}.  In a companion paper \citep[][hereafter
Paper~II]{OL13} we use the local model to analyse the linear
hydrodynamic stability of the laminar flows and find widespread
instability, which requires further investigation in future work.

\section{Large-scale geometry and dynamics of a warped disc}
\label{s:large-scale}

In a thin astrophysical disc, the orbital motion is hypersonic and
fluid elements follow ballistic trajectories to a first approximation.
Around a spherical central mass, these trajectories are Keplerian
orbits, which can have eccentricity and inclination.  A general
Keplerian disc involves smoothly nested orbits of variable
eccentricity and inclination: it is both elliptical and warped.

\begin{figure}
\vskip-0.5cm
\centerline{\epsfysize9cm\epsfbox{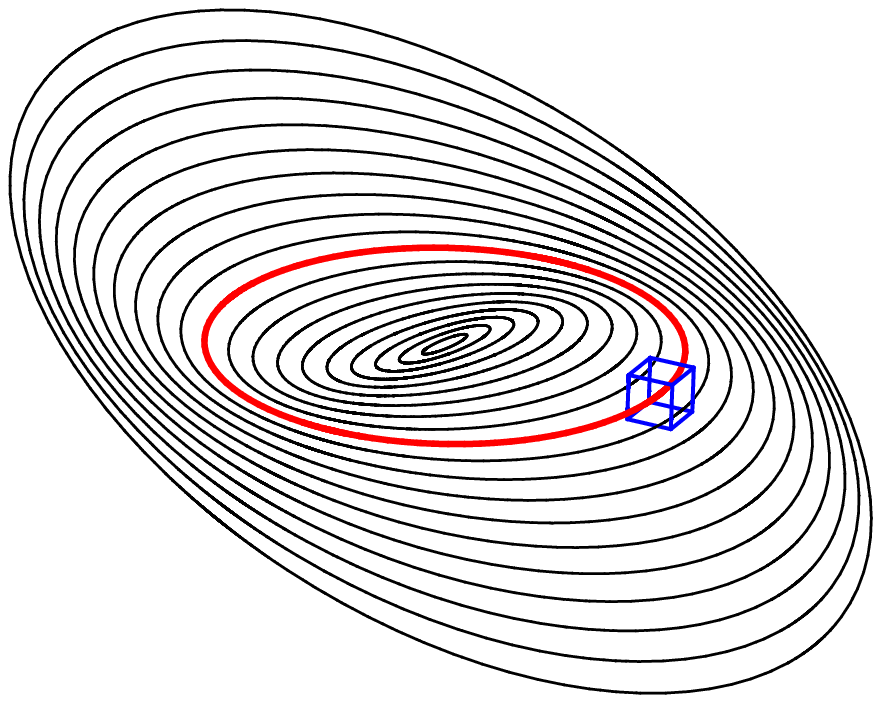}}
\vskip-1cm
\centerline{\epsfysize4cm\epsfbox{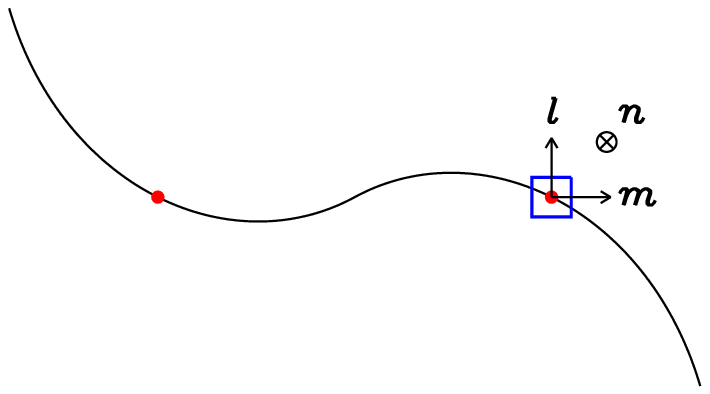}}
\caption{\textit{Top}: Example of an untwisted warped disc, viewed as
  a collection of tilted rings.  As discussed in
  Section~\ref{s:local}, a reference orbit (red circle) is selected
  and used to construct a local Cartesian model of the disc.  The
  local frame (blue box) is centred on a point that follows the
  reference orbit, and therefore experiences a geometry that
  oscillates at the orbital frequency as the local orientation of the
  midplane of the disc tilts back and forth.  The illustrated box
  corresponds to the local frame at orbital phase $0$ according to our
  definitions, and its axes are aligned with the radius-dependent
  basis vectors $\{\bl,\bm,\bn\}$ at this phase only.  In this example
  the warp amplitude at the reference radius is $|\psi|=0.5$.
  \textit{Bottom}: Edge-on view of the warped disc, showing the local
  basis vectors.}
\label{f:warp}
\end{figure}

We consider a spherically symmetric gravitational potential in which
circular orbital motion is possible in any plane containing the centre
of the potential.\footnote{Any small non-spherical component of the
  potential can be considered to provide an external torque on the
  disc.}  The dominant motion in a warped disc is orbital motion in a
plane that varies continuously with radius $r$ and possibly with time
$t$.  In fact the motion need not be circular, but we will not
consider eccentric warped discs in this paper.  We assume that the
time-dependence is slow so that the shape of the warp can be regarded
as fixed on the orbital time-scale.  Therefore the warped disc can be
considered as a continuum of tilted rings (Fig.~\ref{f:warp}, top).
Since astrophysical discs are of non-zero thickness, this structure
can be considered to define the warped midplane of the disc.

The orbital angular velocity of the warped disc is
$\bOmega(r,t)=\Omega(r)\,\bl(r,t)$, where the slowly evolving unit
tilt vector $\bl(r,t)$ is everywhere normal to the local orbital
plane.  The rate of orbital shear is
\begin{equation}
  \bS=r\f{\p\bOmega}{\p r}=r\f{\rmd\Omega}{\rmd r}\,\bl+r\Omega\f{\p\bl}{\p r}=-q\Omega\,\bl+|\psi|\Omega\,\bm,
\label{shear}
\end{equation}
where $q=-\rmd\ln\Omega/\rmd\ln r$ is the usual dimensionless rate of
orbital shear, equal to ${\textstyle\f{3}{2}}$ for Keplerian orbits,
$|\psi|=|\p\bl/\p\ln r|$ is the dimensionless warp amplitude defined
by \citet{1999MNRAS.304..557O}, and $\bm$ is a unit vector parallel to
$\p\bl/\p r$ and therefore orthogonal to $\bl$.  A right-handed
orthonormal triad $\{\bl,\bm,\bn\}$, dependent on $r$ and $t$ but not
on the orbital phase, is completed by $\bn=\bl\times\bm$
(Fig.~\ref{f:warp}, bottom).  Note that none of these vectors is
normal to the disc, in general.

The simplest type of warp is an `untwisted' warp, by which we mean
that, as in Fig.~\ref{f:warp}, the variation of the vectors $\bl$
and~$\bm$ is confined to a plane, while the vector $\bn$ is
perpendicular to that plane and independent of~$r$.  In the language
of celestial mechanics, the longitude of the ascending node is
independent of the semimajor axis.  Although for clarity we choose
untwisted warps for the purposes of illustration, our analysis is
valid for general, twisted warps.  Indeed, the parameter $|\psi|$
measures the local amplitude of the warp whether or not it is twisted.
A local measure of twist would involve the \textit{second} radial
derivative of~$\bl$, for example the triple scalar product of $\bl$,
$\p\bl/\p\ln r$ and $\p^2\bl/\p(\ln r)^2$.  (Some authors, however,
have used the term `twisted disc' as synonymous with `warped disc'.)

\begin{figure}
\centerline{\epsfysize7.5cm\epsfbox{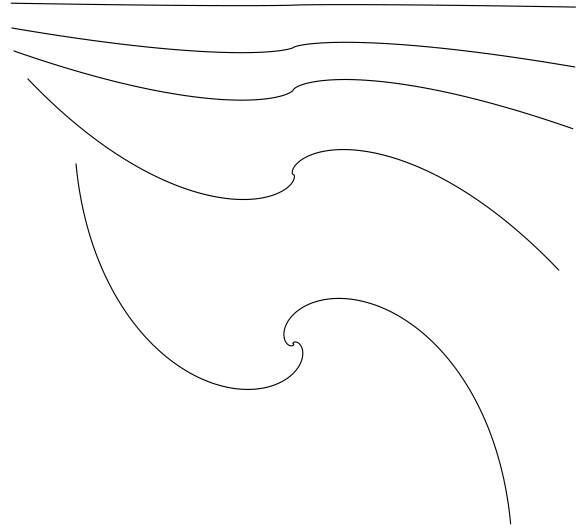}}
\caption{Edge-on view of discs with untwisted warps of constant
  amplitudes $|\psi|=0.01$ (top), $0.1$, $0.2$, $0.5$ and~$1$
  (bottom).  Each curve is a logarithmic spiral.}
\label{f:warps}
\end{figure}

The significance of the parameter $|\psi|$ is illustrated in
Fig.~\ref{f:warps}, where we present an edge-on view of discs with
untwisted warps corresponding to different values of~$|\psi|$.  We
will see in this paper and its companion that a warp of amplitude
$|\psi|=0.01$, which might not be directly observable even with
high-resolution imaging, can nevertheless have important dynamical
consequences.

The large-scale dynamics of a warped disc is governed by the
conservation of mass and angular momentum
\citep{1983MNRAS.202.1181P,1992MNRAS.258..811P}.  The usual aim of a
theory of warped discs is to obtain a system of partial differential
equations in $r$ and $t$ that govern the shape and mass distribution
of the disc.  For a thin disc in which the angular momentum is
dominated by orbital motion, the conservation laws in the absence of
external influences can be written in the one-dimensional form
\begin{equation}
  \f{\p\calM}{\p t}+\f{\p\calF}{\p r}=0,
\label{mass}
\end{equation}
\begin{equation}
  \f{\p}{\p t}(\calM\bh)+\f{\p}{\p r}(\calF\bh+\bcalG)=\mathbf{0},
\label{angmom}
\end{equation}
where $\calM(r,t)$ is the one-dimensional mass density of the disc
(related to the surface density $\Sigma$ through $\calM=2\pi r\Sigma$
and defined such that the mass contained between any two radii is
given by an integral $\int\calM\,\rmd r$ between those radii),
$\calF(r,t)$ is the outward radial flux of mass and
$\bh(r,t)=r^2\bOmega(r,t)=h(r)\bl(r,t)$ is the specific angular
momentum.  The outward radial flux of angular momentum consists of two
parts: $\calF(r,t)\bh(r,t)$ due to advection and $\bcalG(r,t)$ being
an internal torque associated with internal flows and stresses due to
viscosity, magnetic fields, turbulence, self-gravitation, etc.
Subtracting $\bh$ times equation~(\ref{mass}) from
equation~(\ref{angmom}), we obtain the equation governing the shape of
the disc,
\begin{equation}
  \calM h\f{\p\bl}{\p t}+\calF\f{\p\bh}{\p r}+\f{\p\bcalG}{\p r}=\mathbf{0}.
\label{shape}
\end{equation}
A scalar product with the unit vector $\bl$ gives
\begin{equation}
  \calF\f{\rmd h}{\rmd r}+\bl\cdot\f{\p\bcalG}{\p r}=0,
\end{equation}
which determines $\calF$ in terms of~$\bcalG$.  Therefore our task is
to determine the internal torque $\bcalG$.  It is the transport of
angular momentum that drives the evolution of both the shape of the
disc and its mass distribution.  The reason for this is of course
that, within the family of circular orbits, the specific angular
momentum vector determines both the orbital plane and the orbital
radius.

This analysis is easily extended to allow for external forces by
including a source term $\bcalT$, the external torque per unit radius,
on the right-hand side of equation~(\ref{angmom}).

Although the internal torque $\bcalG$ at any radius in a thin disc
involves an integral with respect to the azimuthal angle and is in
this sense a global or large-scale quantity, we will see below that
this integral naturally emerges in the form of a time-average in a
local model that follows the orbital motion.

\section{Local model of a warped disc}
\label{s:local}

\subsection{Geometry and particle dynamics}

A local model can be constructed around any reference point that is
situated on the warped midplane of the disc and moves in a circular
orbit of radius $r_0$, where the orbital angular velocity is
$\Omega_0=\Omega(r_0)$.  A Cartesian coordinate system is set up with
its origin at the moving reference point and with axes that rotate
with the orbital motion, so that the $x$, $y$ and~$z$ directions are
always radial, azimuthal and vertical.  This is the standard
construction of a local model, as used in the shearing sheet or
shearing box.

The effective potential in this frame is the sum of the gravitational
potential and the centrifugal potential arising from the rotation of
the frame.  When the effective potential is expanded in a Taylor
series about the reference point and the unimportant constant term is
neglected, the dominant terms are
$-q\Omega_0^2x^2+\half\Omega_0^2z^2$.  The equations of particle
dynamics in this model are therefore
\begin{equation}
  \ddot x-2\Omega_0\dot y=2q\Omega_0^2x,
\label{hillx}
\end{equation}
\begin{equation}
  \ddot y+2\Omega_0\dot x=0,
\end{equation}
\begin{equation}
  \ddot z=-\Omega_0^2z,
\label{hillz}
\end{equation}
which reduce to Hill's equations (without a satellite) in the
Keplerian case $q={\textstyle\f{3}{2}}$.

\begin{figure}
\centerline{\epsfysize5cm\epsfbox{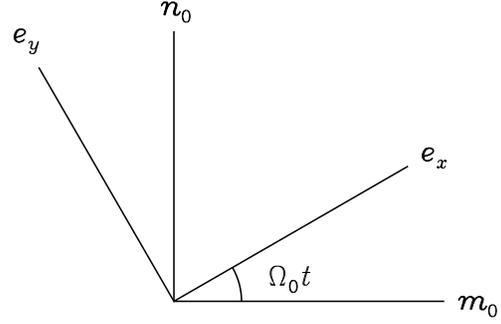}}
\caption{Time-dependent relation between the rotating horizontal basis
  vectors $\be_x$ and $\be_y$ of the local model and the non-rotating
  basis vectors $\bm_0$ and $\bn_0$ associated with the geometry of
  the warp at the reference radius.  The vertical basis vectors
  $\be_z$ and $\bl_0$ both point out of the page.}
\label{f:basis}
\end{figure}

As stated above, we assume that the geometry of the warped disc can be
regarded as fixed on the orbital time-scale relevant to the local
model, although we discuss this assumption further in
Section~\ref{s:role} below.  The basis vectors associated with the
warp at the reference radius, $\{\bl_0,\bm_0,\bn_0\}$, are therefore
non-rotating, while the basis $\{\be_x,\be_y,\be_z\}$ rotates as the
reference point traverses its orbit.  Without loss of generality, we
choose the orbital phase of the reference point, or the origin of
time, such that $\bm_0$ is in the radial direction at $t=0$ (as in
Fig.~\ref{f:warp}).  Then the basis vectors are related by
\begin{equation}
  \bl_0=\be_z,\qquad
  \bm_0+\rmi\bn_0=(\be_x+\rmi\be_y)\rme^{\rmi\Omega_0t}
\label{bases}
\end{equation}
(Fig.~\ref{f:basis}).  While the rate of orbital shear is stationary
in a non-rotating frame, in the local frame it rotates in a negative
sense according to
\begin{equation}
  \bS_0(t)=-q\Omega_0\,\be_z+|\psi|\Omega_0[\be_x\cos(\Omega_0t)-\be_y\sin(\Omega_0t)],
\end{equation}
which follows from equations~(\ref{shear}) and~(\ref{bases}).  In
Fig.~\ref{f:warp} the box can be thought of as following the red
reference orbit, experiencing a local geometry that oscillates at the
orbital frequency as the local orientation of the midplane of the disc
tilts back and forth.

The local representation of circular orbital motion in the $xy$ plane
is the linear motion $\dot y=-q\Omega_0x$ with $x=\cst$.  To this can
be added a free vertical oscillation
$z=\real(Z\,\rme^{-\rmi\Omega_0t})$.  The amplitude of the complex
number $Z$ is related to the (small) inclination of the orbit with
respect to the $xy$ plane, while the phase of~$Z$ is related to the
longitude of the ascending node; the complex tilt variable (used by
\citealt{1983MNRAS.202.1181P} and many other authors), measured with
respect to the $xy$ plane, would be $W=-Z/r_0$.

An orbital motion with a smooth warp can be represented locally by
allowing $Z$ to vary linearly with $x$.  Since the reference point is
on the warped midplane of the disc, $Z$ vanishes at $x=0$.  With the
choice of orbital phase explained above, we have $Z=-|\psi|x$, and so
$z=-|\psi|\cos(\Omega_0t)x$.  This motion corresponds to the velocity
field
\begin{equation}
  \dot\bx=\bu=-q\Omega_0x\,\be_y+|\psi|\Omega_0\sin(\Omega_0t)x\,\be_z.
\label{orbital}
\end{equation}
Thus, while a particle at the origin remains there (corresponding to
the reference orbit), a particle with a positive value of $x$ lags
behind and also oscillates up and down at the orbital frequency
(corresponding to an orbit slightly larger than the reference orbit).

Later we will need to calculate the internal torque in the local
model.  To prepare the way for this, we consider here the specific
angular momentum, which in the local model can be regarded as
\begin{equation}
  r_0(\dot y+2\Omega_0x)\be_z-r_0\dot z\,\be_y-r_0\Omega_0z\,\be_x.
\label{angmom_local}
\end{equation}
The first two vectors involve non-radial motions combined with the
long radial lever arm, while the third vector involves the large
azimuthal motion combined with a vertical lever arm.  (The
$2\Omega_0x$ term represents the contribution to the azimuthal motion
from the rotation of the frame of reference.)  If we had also included
the large azimuthal motion combined with the long radial lever arm, we
would have obtained an additional term $r_0^2\Omega_0\,\be_z=\bh_0$,
which is constant and large compared to the terms considered here.
For a general particle motion governed by equations
(\ref{hillx})--(\ref{hillz}), it can be verified that the
vector~(\ref{angmom_local}) is constant in a non-rotating frame. Its
components in the local frame, however, are not constant.

For the orbital motion~(\ref{orbital}) associated with a warped disc,
the specific angular momentum evaluates to
\begin{equation}
  r_0(\bS_0+2\bOmega_0)x=\left(\f{\p\bh}{\p r}\right)_0x,
\label{angmom_orbital}
\end{equation}
and therefore agrees with a linear local approximation to the orbital
angular momentum.  Again, if we had also included the large azimuthal
motion combined with the long radial lever arm, we would have obtained
an additional term $\bh_0$, which is constant and large compared to
the terms considered here.

\subsection{Fluid dynamics}

So far we have considered the motion of test particles.  The local
hydrodynamic solutions for a warped disc are more complicated than the
particle motion~(\ref{orbital}) because of pressure gradients.  These
solutions generally require numerical calculations and are most easily
obtained by introducing a coordinate transformation that accounts for
the warped orbital motion.  Before doing so we establish the equations
to be solved.

The basic equations for an ideal fluid in the local model (neglecting
magnetic fields and self-gravity) are
\begin{equation}
  \rmD u_x-2\Omega_0u_y=2q\Omega_0^2x-\f{1}{\rho}\p_xp,
\end{equation}
\begin{equation}
  \rmD u_y+2\Omega_0u_x=-\f{1}{\rho}\p_yp,
\end{equation}
\begin{equation}
  \rmD u_z=-\Omega_0^2z-\f{1}{\rho}\p_zp,
\end{equation}
\begin{equation}
  \rmD\rho=-\rho(\p_xu_x+\p_yu_y+\p_zu_z),
\end{equation}
where
\begin{equation}
  \rmD=\p_t+u_x\p_x+u_y\p_y+u_z\p_z
\label{d}
\end{equation}
is the Lagrangian derivative, $\bu$ is the velocity, $\rho$ is the
density and $p$ is the pressure.  These are the standard equations for
hydrodynamics in the shearing sheet or box
\citep[e.g.][]{1995ApJ...440..742H}.  For simplicity, we consider an
isothermal gas for which
\begin{equation}
  p=c_\rms^2\rho,
\end{equation}
where $c_\rms=\cst$ is the isothermal sound speed.  In terms of the
pseudo-enthalpy
\begin{equation}
  h=c_\rms^2\ln\rho+\cst,
\end{equation}
we then have
\begin{equation}
  \rmD u_x-2\Omega_0u_y=2q\Omega_0^2x-\p_xh,
\label{dux}
\end{equation}
\begin{equation}
  \rmD u_y+2\Omega_0u_x=-\p_yh,
\end{equation}
\begin{equation}
  \rmD u_z=-\Omega_0^2z-\p_zh,
\label{duz}
\end{equation}
\begin{equation}
  \rmD h=-c_\rms^2(\p_xu_x+\p_yu_y+\p_zu_z).
\label{dh1}
\end{equation}
The alternative equations for adiabatic flow are developed in
Appendix~\ref{s:adiabatic}.

Note that our local model correctly includes the vertical
gravitational acceleration $-\Omega_0^2z$ deriving from a spherically
symmetric potential.  Without this term the model would be incapable
of representing a warped disc and the dynamics described in this paper
would not occur.

If the disc is unwarped, the orbital plane everywhere coincides with
the $xy$ plane and the local representation of circular orbital motion
is $\bu=-q\Omega_0x\,\be_y$.  Equations (\ref{dux})--(\ref{dh1}) are
then satisfied when hydrostatic equilibrium, $\p_zh=-\Omega_0^2z$, is
also imposed.

If the disc is warped at the reference radius, then we might expect
the velocity field~(\ref{orbital}) corresponding to the warped orbital
motion to satisfy the hydrodynamic equations (\ref{dux})--(\ref{dh1}),
together with some version of hydrostatic equilibrium.  However, this
is not the case; the problem lies with equation~(\ref{duz}).  If
hydrostatic equilibrium, $\p_zh=-\Omega_0^2z$, is imposed, then there
is an unbalanced, $x$-dependent acceleration on the left-hand side of
equation~(\ref{duz}).  If the pressure is allowed to vary with $x$ to
compensate for this term, then equation~(\ref{dux}) is disrupted.  The
actual hydrodynamic solutions must be more complicated than
equation~(\ref{orbital}) in order to account for the interplay of
these additional forces.  The simplest (laminar) versions of these
solutions are derived in Section~\ref{s:laminar} below.

\subsection{Warped shearing coordinates}
\label{s:coordinates}

We now adapt the local model to incorporate the oscillating local
geometry of the warp explicitly.  This will allow us to find the
simplest hydrodynamic states and to formulate a theoretical and
computational model for the further study of warped discs.

\begin{figure*}
\centerline{\epsfysize7.5cm\epsfbox{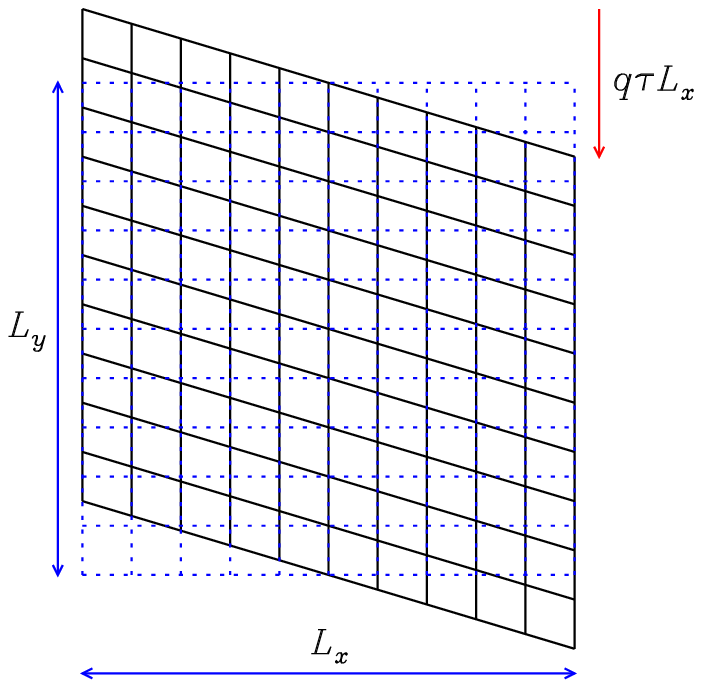}\epsfysize7.5cm\epsfbox{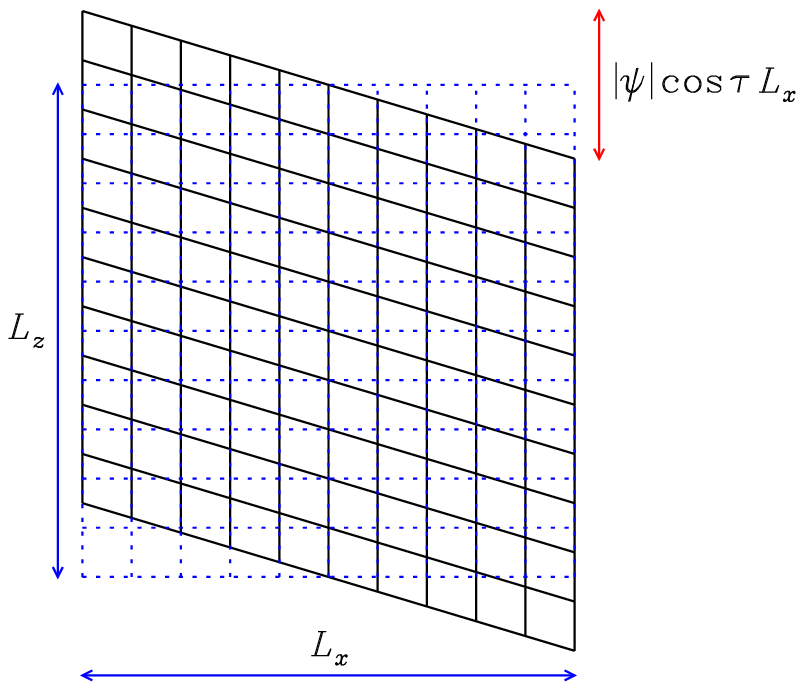}}
\caption{Illustration of warped shearing coordinates.  The dotted blue
  grid represents the Cartesian coordinates of the standard local
  approximation or shearing box, while the solid black grid represents
  the warped shearing coordinates.  The shear in the $xy$ plane (left)
  increases linearly with time, although in a computational model the
  grid should be remapped periodically by resetting the origin of
  time.  The shear in the $xz$ plane (right) oscillates at the orbital
  frequency and is proportional to the amplitude of the warp.}
\label{f:grid}
\end{figure*}

We introduce new coordinates
\begin{equation}
  t'=t,
\end{equation}
\begin{equation}
  x'=x,
\label{xp}
\end{equation}
\begin{equation}
  y'=y+q\Omega_0tx,
\label{yp}
\end{equation}
\begin{equation}
  z'=z+|\psi|\cos(\Omega_0t)x
\end{equation}
that follow the warped orbital motion: a particle on an inclined
circular orbit would have constant $x'$, $y'$ and~$z'$.  We define the
orbital phase
\begin{equation}
  \tau=\Omega_0t'=\Omega_0t.
\end{equation}
Partial derivatives transform
according to
\begin{equation}
  \p_t=\p_t'+q\Omega_0x\p_y'-|\psi|\Omega_0\sin\tau\,x\p_z',
\end{equation}
\begin{equation}
  \p_x=\p_x'+q\tau\,\p_y'+|\psi|\cos\tau\,\p_z',
\end{equation}
\begin{equation}
  \p_y=\p_y',
\end{equation}
\begin{equation}
  \p_z=\p_z',
\end{equation}
so that
\begin{equation}
  \rmD=\p_t'+v_x\p_x'+(v_y+q\tau v_x)\p_y'+(v_z+|\psi|\cos\tau\,v_x)\p_z',
\end{equation}
where
\begin{equation}
  v_x=u_x,
\end{equation}
\begin{equation}
  v_y=u_y+q\Omega_0x,
\end{equation}
\begin{equation}
  v_z=u_z-|\psi|\Omega_0\sin\tau\,x
\end{equation}
are the relative velocity components.  Thus, if $\bv=\mathbf{0}$
(which is \textit{not} a solution of the equations), the fluid follows
the prescribed warped orbital motion and the velocity $\bu$, which
corresponds to the rate of change of the Cartesian coordinates
$(x,y,z)$, is non-zero.  (We should not call $\bu$ the absolute
velocity, however, because it is measured in a rotating frame of
reference.)  We continue to refer vector components to the Cartesian
basis $\{\be_x,\be_y,\be_z\}$.  The grid of \textit{warped shearing
  coordinates} is illustrated in Fig.~\ref{f:grid}.

The hydrodynamic equations are therefore transformed into
\begin{eqnarray}
  \rmD v_x-2\Omega_0v_y=-(\p_x'+q\tau\,\p_y'+|\psi|\cos\tau\,\p_z')h,
\label{dvx}
\end{eqnarray}
\begin{equation}
  \rmD v_y+(2-q)\Omega_0v_x=-\p_y'h,
\end{equation}
\begin{equation}
  \rmD v_z+|\psi|\Omega_0\sin\tau\,v_x=-\Omega_0^2z'-\p_z'h,
\label{dvz}
\end{equation}
\begin{equation}
  \rmD h=-c_\rms^2[(\p_x'+q\tau\,\p_y'+|\psi|\cos\tau\,\p_z')v_x+\p_y'v_y+\p_z'v_z].
\label{dh}
\end{equation}
An alternative form of equation~(\ref{dh}), in which the conservation
of mass is manifest, is
\begin{eqnarray}
  \lefteqn{\p_t'\rho+\p_x'(\rho v_x)+\p_y'[\rho(v_y+q\tau v_x)]}&\nonumber\\
  &&+\p_z'[\rho(v_z+|\psi|\cos\tau\,v_x)]=0.
\end{eqnarray}
These equations contain $\tau$ explicitly but not $x'$ or $y'$.  This
shows that the local model of a warped disc is horizontally
homogeneous.  As in the standard shearing sheet, every point in the
$x'y'$ plane is equivalent, if allowance is made for an appropriate
change of frame of reference.

\begin{figure}
\bigskip
\centerline{\epsfysize3cm\epsfbox{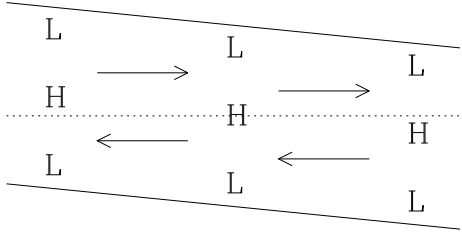}}
\bigskip
\caption{Regions of high (H) and low (L) pressure in the local
  representation of a warped disc, showing how the hydrostatic
  vertical pressure gradient leads to a horizontal force that is
  unbalanced by gravity (arrows).  The $xz$ plane (of unwarped
  coordinates) is depicted at $t=0$, with the dotted line
  corresponding to $z=0$.  These horizontal pressure gradients
  oscillate at the orbital frequency as the local geometry tilts back
  and forth, giving rise to oscillatory planar motions.}
\label{f:unbalanced}
\end{figure}

We can see from these equations that $\bv=\mathbf{0}$ is not a
solution in the presence of a warp.  Equation~(\ref{dvz}) would
require hydrostatic equilibrium in the form $\p_z'h=-\Omega_0^2z'$,
but this would provide an unbalanced horizontal force in
equation~(\ref{dvx}), as illustrated in Fig.~\ref{f:unbalanced}.

The total energy equation associated with equations
(\ref{dvx})--(\ref{dh}) is
\begin{eqnarray}
  \lefteqn{\p_t'(\rho\mathcal{E})+\p_x'[(\rho\mathcal{E}+p)v_x]+\p_y'[(\rho\mathcal{E}+p)(v_y+q\tau v_x)]}&\nonumber\\
  &&+\p_z'[(\rho\mathcal{E}+p)(v_z+|\psi|\cos\tau\,v_x)]\nonumber\\
  &&=\rho v_x(q\Omega_0v_y-|\psi|\Omega_0\sin\tau\,v_z+|\psi|\Omega_0\cos\tau\,\Omega_0z'),
\label{total_energy}
\end{eqnarray}
where
\begin{equation}
  \mathcal{E}=\f{1}{2}(v_x^2+v_y^2+v_z^2)+h+\f{1}{2}\Omega_0^2z'^2.
\end{equation}
We will interpret the source term on the right-hand side of
equation~(\ref{total_energy}) below.

\subsection{Fluxes of mass and angular momentum}
\label{s:fluxes}

In order to connect the local model with the large-scale dynamics of
the warped disc described in Section~\ref{s:large-scale}, we consider
the outward radial fluxes of mass and angular momentum.  The mass flux
at the reference radius $r_0$ is
\begin{equation}
  \calF_0=\int\!\!\int_0^{2\pi}\langle\rho u_x\rangle_\rmh\,r_0\,\rmd\tau\,\rmd z'
\label{calF0}
\end{equation}
(in which $u_x$ can be replaced by $v_x$), where the notation
$\langle\cdot\rangle_\rmh$ denotes local horizontal averaging over the
coordinates $x'$ and~$y'$ (where necessary), and the $z'$ integral is
over the entire vertical extent of the disc.  The $\tau$ integral,
which in the local model is with respect to time, can be interpreted
as an integral with respect to azimuth as the reference point
traverses its orbit.  Similarly, the angular-momentum flux in the
local model is (cf.\ equations~\ref{angmom_local}
and~\ref{angmom_orbital})
\begin{eqnarray}
  \lefteqn{\calF_0\left(\f{\p\bh}{\p r}\right)_0x+\bcalG_0=\int\!\!\int_0^{2\pi}\big\langle\rho u_x[r_0(u_y+2\Omega_0x)\,\be_z-r_0u_z\,\be_y}&\nonumber\\
  &&\qquad\qquad\qquad\qquad\qquad-r_0\Omega_0z\,\be_x]\big\rangle_\rmh\,r_0\,\rmd\tau\,\rmd z'.
\end{eqnarray}
As in equation~(\ref{angmom_local}), the first two vectors inside the
square brackets, contributing to the specific angular momentum of the
fluid, involve non-radial motions combined with the long radial lever
arm, while the third expression involves the large azimuthal motion
combined with a vertical lever arm; again, we choose not to include
the larger constant term involving the large azimuthal motion combined
with the long radial lever arm, which generates the larger flux
$\calF_0\bh_0$.  Writing $\bu$ and~$z$ in terms of~$\bv$ and~$z'$, and
using the definition~(\ref{calF0}) of $\calF_0$, we find that the
terms proportional to $x$ on each side of the equation cancel, leaving
\begin{equation}
  \bcalG_0=\int\!\!\int_0^{2\pi}\langle\bg_0\rangle_\rmh\,r_0\,\rmd\tau\,\rmd z',
\label{calG0}
\end{equation}
where
\begin{equation}
  \bg_0=\rho v_x(r_0v_y\,\be_z-r_0v_z\,\be_y-r_0\Omega_0z'\,\be_x)
\label{g0}
\end{equation}
is the internal torque per unit area.  We then recognize the
right-hand side of equation~(\ref{total_energy}) as the scalar product
$-(\bg_0\cdot\bS_0)/r_0$, which is the rate at which energy is
extracted, per unit volume, by the torque acting on the orbital shear.

So far we have considered inviscid hydrodynamics.  More generally,
there may be shear stresses in the fluid resulting from viscosity or
magnetic fields.  In the presence of a symmetric stress tensor
$\mathbf{T}$, which can also represent the effects of turbulence or
self-gravitation, the torque per unit area is given by the more
general expression
\begin{equation}
  \f{\bg_0}{r_0}=\rho v_x(v_y\,\be_z-v_z\,\be_y-\Omega_0z'\,\be_x)-T_{xy}\,\be_z+T_{xz}\,\be_y.
\label{g02}
\end{equation}
In the case of a viscous stress, for example,
\begin{equation}
  T_{xy}=\mu[-q\Omega_0+(\p_x'+q\tau\,\p_y'+|\psi|\cos\tau\,\p_z')v_y+\p_y'v_x]
\end{equation}
and
\begin{equation}
  T_{xz}=\mu[|\psi|\Omega_0\sin\tau+(\p_x'+q\tau\,\p_y'+|\psi|\cos\tau\,\p_z')v_z+\p_z'v_x]
\end{equation}
are the relevant components, where $\mu$ is the dynamic viscosity.

Since $\bg_0\cdot\bl_0=g_{0z}$, the vertical component of the torque
is given by
\begin{equation}
  \f{\bcalG_0\cdot\bl_0}{2\pi r_0^2}=\bigg\langle\int(\rho v_xv_y-T_{xy})\,\rmd z'\bigg\rangle_{\!\!\rmh,\tau},
\end{equation}
where the notation $\langle\cdot\rangle_{\rmh,\tau}$ denotes
horizontal averaging over the coordinates $x'$ and~$y'$ (where
necessary) and averaging over the orbital phase $\tau$.  This is
easily understood as involving the radial transport of azimuthal
momentum (or vertical angular momentum) by Reynolds and viscous (or
other) stresses.

In computing the horizontal components of the torque we must bear in
mind that the basis vectors $\be_x$ and $\be_y$ depend on orbital
phase.  We make use of a complex notation and equation~(\ref{bases}).
Thus
\begin{equation}
  \bg_0\cdot(\bm_0+\rmi\bn_0)=(g_{0x}+\rmi g_{0y})\rme^{\rmi\tau}
\end{equation}
and so
\begin{eqnarray}
  \lefteqn{\f{\bcalG_0\cdot(\bm_0+\rmi\bn_0)}{2\pi r_0^2}=}&\nonumber\\
  &&\qquad\bigg\langle\rme^{\rmi\tau}\int[\rho v_x(-\Omega_0z'-\rmi v_z)+\rmi T_{xz}]\,\rmd z'\bigg\rangle_{\!\!\rmh,\tau}.
\end{eqnarray}
The horizontal components of the torque are conveniently encoded in
the real and imaginary parts of this expression.  Not surprisingly, it
involves the radial transport of vertical momentum (or horizontal
angular momentum).

It is natural to represent the torque for an isothermal disc in the
form \citep[cf.][]{1999MNRAS.304..557O}
\begin{eqnarray}
  \bcalG&=&-2\pi\Sigma c_\rms^2r^2(Q_1\,\bl+Q_2|\psi|\,\bm+Q_3|\psi|\,\bn)\nonumber\\
  &=&-2\pi\Sigma c_\rms^2r^2\left(Q_1\,\bl+Q_2r\f{\p\bl}{\p r}+Q_3r\,\bl\times\f{\p\bl}{\p r}\right),
\label{torque}
\end{eqnarray}
where $Q_1$, $Q_2$
and $Q_3$ are dimensionless coefficients and
\begin{equation}
  \Sigma=\bigg\langle\int\rho\,\rmd z'\bigg\rangle_{\!\!\rmh}
\end{equation}
is the horizontally averaged surface density of the disc, which is
independent of~$\tau$ by virtue of mass conservation.  This
representation is just a projection of the torque on to the local
basis defined by the geometry of the disc.\footnote{In the absence of
  a warp, $|\psi|$ vanishes and $\bm$ and $\bn$ are undefined, but in
  this case we expect $\bcalG$ to be parallel to $\bl$ by symmetry.}
Comparison of the vertical component of the torque shows that
\begin{equation}
  -Q_1\Sigma c_\rms^2=\bigg\langle\int(\rho v_xv_y-T_{xy})\,\rmd z'\bigg\rangle_{\!\!\rmh,\tau},
\end{equation}
while comparison of the horizontal components shows that
\begin{equation}
  -Q_4|\psi|\Sigma c_\rms^2=\bigg\langle\rme^{\rmi\tau}\int[\rho v_x(-\Omega_0z'-\rmi v_z)+\rmi T_{xz}]\,\rmd z'\bigg\rangle_{\!\!\rmh,\tau},
\end{equation}
where $Q_4=Q_2+\rmi Q_3$ is a useful combination.  Roughly speaking,
the $Q_1$ term is similar to the usual torque in an accretion disc,
and (if negative) causes the mass distribution to evolve diffusively,
while the $Q_2$ term (if positive) causes a diffusion of the warp and
the $Q_3$ term causes a dispersive wavelike propagation of the warp.

The correspondence between these dimensionless torque coefficients and
the viscosity coefficients $\nu_1$ and $\nu_2$ of
\citet{1992MNRAS.258..811P} (deriving from the `naive approach' in
Section~2 of \citealt{1983MNRAS.202.1181P}) is
\begin{equation}
  Q_1c_\rms^2=-q\nu_1\Omega,\qquad
  Q_2c_\rms^2=\half\nu_2\Omega,
\end{equation}
while $Q_3$ has no counterpart in that description.  (It would be
potentially misleading to associate $Q_3$ with a `viscosity'
coefficient $\nu_3$ because the $Q_3$ term is non-dissipative.
However, the combination $Q_4=Q_2+\rmi Q_3$ does emerge naturally as a
complex diffusion coefficient, especially in linear theory.)

Since the shape of the disc is defined in the local model only by the
dimensionless warp amplitude $|\psi|$, it is natural to expect the
dimensionless torque coefficients $Q_1$, $Q_2$ and $Q_3$ to emerge as
functions of $|\psi|$, as well as any other relevant dimensionless
parameters such as $\alpha$.  We will see in Section~\ref{s:torques}
how this works for laminar flows, but in the presence of instability
and turbulence these functions must be determined by means of
numerical simulations.

\subsection{Computational considerations}

The hydrodynamic equations (\ref{dvx})--(\ref{dh}) or their
equivalents can be solved numerically by standard methods, for example
by using finite differences on a regular grid in the coordinates
$(x',y',z')$ and imposing periodic boundary conditions in $x'$ and
$y'$ and some other appropriate boundary conditions in $z'$.  The
additional terms proportional to $|\psi|$ would of course require some
rewriting of existing codes.  Note that the primed coordinate grid
undergoes an inexorable shearing, even in the absence of a warp.  The
coordinates should therefore be remapped periodically to avoid
excessive distortion of the grid.  This procedure is equivalent to
periodically resetting the arbitrary origin of time, and can be done
without interpolation if the remapping frequency and the aspect ratio
of the grid are chosen correctly.  In the case of a warped disc with
its oscillating local geometry, it may be convenient to remap once per
orbital period, for example by letting $\tau$ run from $-\pi$ to $\pi$
repeatedly, giving rise to a periodic dynamical system.

In the standard shearing box for an unwarped disc, there are in fact
two different methods of solving the hydrodynamic equations.  The more
common method is to discretize the equations (with time-independent
coefficients) in a cuboidal domain in unsheared coordinates $(x,y,z)$
and to apply (time-dependent) modified periodic boundary conditions.
The other method is to discretize the equations (with time-dependent
coefficients) in a cuboidal domain in sheared coordinates $(x',y',z')$
(with periodic remapping of the grid) and to apply (time-independent)
periodic boundary conditions.  For a warped disc, only the second
approach is possible: the equations must be discretized in primed
coordinates, because a fixed cuboidal domain in unprimed coordinates
cannot represent the regions above and below the warped midplane of
the disc in a way that allows the radial boundaries to be identified
(cf.\ Fig.~\ref{f:unbalanced}).  This limitation is related to the
fact that vertical gravity is essential to the description of a warped
disc, and the $z$ direction cannot be regarded as periodic.  When
supplied with periodic boundary conditions in $x'$ and~$y'$, our local
model takes the form of a \textit{warped shearing box}.

\section{Laminar flows}
\label{s:laminar}

\subsection{Separation of variables}
\label{s:separation}

The simplest solutions of equations (\ref{dvx})--(\ref{dh}) are in
fact independent of~$x'$ and~$y'$ and $2\pi$-periodic in $\tau$: they
are horizontally uniform and oscillate at the orbital frequency.  (In
a global context, such local solutions correspond to the situation in
which the warped disc has a structure that varies on a horizontal
length-scale much larger than the thickness of the disc, and on a
time-scale much longer than the orbital time-scale.)  These laminar
flows satisfy
\begin{equation}
  \rmD v_x-2\Omega_0v_y=-|\psi|\cos\tau\,\p_z'h,
\end{equation}
\begin{equation}
  \rmD v_y+(2-q)\Omega_0v_x=0,
\end{equation}
\begin{equation}
  \rmD v_z+|\psi|\Omega_0\sin\tau\,v_x=-\Omega_0^2z'-\p_z'h,
\end{equation}
\begin{equation}
  \rmD h=-c_\rms^2(|\psi|\cos\tau\,\p_z'v_x+\p_z'v_z),
\end{equation}
with
\begin{equation}
  \rmD=\p_t'+(v_z+|\psi|\cos\tau\,v_x)\p_z'.
\end{equation}
A nonlinear separation of variables is then possible, with
\begin{equation}
  v_x(z',t')=u(\tau)\Omega_0z',
\label{vx}
\end{equation}
\begin{equation}
  v_y(z',t')=v(\tau)\Omega_0z',
\end{equation}
\begin{equation}
  v_z(z',t')=w(\tau)\Omega_0z',
\end{equation}
\begin{equation}
  h(z',t')=c_\rms^2f(\tau)-\half\Omega_0^2z'^2g(\tau),
\label{h}
\end{equation}
where $u$, $v$, $w$, $f$ and~$g$ are all dimensionless.  Thus
\begin{equation}
  \rmd_\tau u+(w+|\psi|\cos\tau\,u)u-2v=|\psi|\cos\tau\,g,
\end{equation}
\begin{equation}
  \rmd_\tau v+(w+|\psi|\cos\tau\,u)v+(2-q)u=0,
\end{equation}
\begin{equation}
  \rmd_\tau w+(w+|\psi|\cos\tau\,u)w+|\psi|\sin\tau\,u=g-1,
\end{equation}
\begin{equation}
  \rmd_\tau f=-(w+|\psi|\cos\tau\,u),
\end{equation}
\begin{equation}
  \rmd_\tau g=-2(w+|\psi|\cos\tau\,u)g,
\end{equation}
where $\rmd_\tau$ denotes the ordinary derivative $\rmd/\rmd\tau$.

When a dynamic shear viscosity $\mu=\alpha p/\Omega_0$ and a dynamic
bulk viscosity $\mu_\rmb=\alpha_\rmb p/\Omega_0$ are included, where
$\alpha$ and $\alpha_\rmb$ are constant dimensionless coefficients
(thus providing a kinematic shear viscosity $\nu=\alpha
c_\rms^2/\Omega_0$, etc.), these equations become
\begin{eqnarray}
  \lefteqn{\rmd_\tau u+(w+|\psi|\cos\tau\,u)u-2v=|\psi|\cos\tau\,g}&\nonumber\\
  &&-(\alpha_\rmb+\third\alpha)|\psi|\cos\tau\,g(w+|\psi|\cos\tau\,u)\nonumber\\
  &&-\alpha g[|\psi|\sin\tau+(1+|\psi|^2\cos^2\tau)u],
\label{du}
\end{eqnarray}
\begin{eqnarray}
  \lefteqn{\rmd_\tau v+(w+|\psi|\cos\tau\,u)v+(2-q)u}&\nonumber\\
  &&=-\alpha g[-q|\psi|\cos\tau+(1+|\psi|^2\cos^2\tau)v],
\end{eqnarray}
\begin{eqnarray}
  \lefteqn{\rmd_\tau w+(w+|\psi|\cos\tau\,u)w+|\psi|\sin\tau\,u=g-1}&\nonumber\\
  &&-(\alpha_\rmb+\third\alpha)g(w+|\psi|\cos\tau\,u)\nonumber\\
  &&-\alpha g[|\psi|^2\sin\tau\cos\tau+(1+|\psi|^2\cos^2\tau)w],
\end{eqnarray}
\begin{equation}
  \rmd_\tau f=-(w+|\psi|\cos\tau\,u),
\label{df}
\end{equation}
\begin{equation}
  \rmd_\tau g=-2(w+|\psi|\cos\tau\,u)g.
\label{dg}
\end{equation}

Our numerical treatment of these ordinary differential equations in
Section~\ref{s:numerical_laminar} below provides us with a family of
`exact' nonlinear solutions representing the laminar flows in a warped
disc.  The same laminar flows, combined with the warping motion, can
alternatively be regarded as nonlinear solutions of equations
(\ref{dux})--(\ref{dh1}) in the variables of the standard shearing
sheet, although they do not then appear to be horizontally uniform and
it would not be obvious how to obtain them without making the
transformation to warped coordinates.  For example, the vertical
velocity would be
\begin{equation}
  u_z=|\psi|\Omega_0\sin\tau\,x+w(\tau)\Omega_0(z+|\psi|\cos\tau\,x).
\end{equation}

The differential equations for $f$ and~$g$ are related in such a way
that the surface density of the disc, which is proportional to
$\rme^fg^{-1/2}$, is independent of~$\tau$.  The total energy equation
for the laminar flows has the form
\begin{eqnarray}
  \lefteqn{\rmd_\tau\left[\f{1}{2g}(u^2+v^2+w^2+g\ln g+1)\right]}&\nonumber\\
  &&=\f{1}{g}(quv-|\psi|\sin\tau\,uw+|\psi|\cos\tau\,u)\nonumber\\
  &&-\alpha|\psi|(\sin\tau\,u-q\cos\tau\,v+|\psi|\sin\tau\cos\tau\,w)\nonumber\\
  &&-\alpha(1+|\psi|^2\cos^2\tau)(u^2+v^2+w^2)\nonumber\\
  &&-(\alpha_\rmb+\third\alpha)(w+|\psi|\cos\tau\,u)^2.
\end{eqnarray}
The first expression on the right-hand side is equivalent to the
right-hand side of equation~(\ref{total_energy}) and corresponds to
the extraction of energy by the torque acting on the orbital shear.
The third and fourth expressions are negative definite and correspond
to viscous damping of the flow.  The second expression is proportional
to viscosity but is not negative definite; its interpretation is not
straightforward.

In Appendix~\ref{s:adiabatic} we give the corresponding equations for
adiabatic flow.  They are exactly equivalent to those derived by
\citet{1999MNRAS.304..557O} in a different way, by writing the
hydrodynamic equations in a warped spherical polar coordinate system
that follows the global distortion of the warped midplane, and
carrying out an asymptotic expansion of the solution in the limit of a
thin disc.  In that paper the orbital phase $\tau$ is directly related
to the azimuthal angle $\phi$.  Here we have not explicitly made use
of an asymptotic expansion, but have derived the equations
consistently within the context of the local approximation.

\subsection{Horizontal and vertical oscillators}

The physical interpretation of the laminar flows is assisted by
changing to the variables
\begin{equation}
  (U,V,W)=(u,v,w)H,\qquad
  H=g^{-1/2}.
\end{equation}
Since the density and pressure then have the form
\begin{equation}
  \rho\propto p\propto\exp\left\{f(\tau)-\f{1}{2}\left[\f{\Omega_0z'}{c_\rms H(\tau)}\right]^2\right\},
\end{equation}
it can be seen that $H(\tau)$ is the scale-height of the disc in units
of the scale-height $c_\rms/\Omega_0$ of an equilibrium unwarped disc.
The variables $(U,V,W)$ are therefore more like momenta, or vertically
integrated velocities.

In the inviscid case we then have
\begin{equation}
  \rmd_\tau U-2V=|\psi|\cos\tau\,H^{-1},
\label{duu}
\end{equation}
\begin{equation}
  \rmd_\tau V+(2-q)U=0,
\label{dvv}
\end{equation}
\begin{equation}
  \rmd_\tau W+|\psi|\sin\tau\,U=H^{-1}-H,
\label{dww}
\end{equation}
\begin{equation}
  \rmd_\tau H=W+|\psi|\cos\tau\,U.
\label{dhh}
\end{equation}
In the absence of a warp ($|\psi|=0$), equations~(\ref{duu})
and~(\ref{dvv}) describe a linear epicyclic oscillator.  They can be
combined to give $\rmd_\tau^2V=-2(2-q)V$, or the same equation for
$U$.  Equations~(\ref{dww}) and~(\ref{dhh}) describe a nonlinear
vertical oscillator, the `breathing mode' of the disc.  They can be
combined to give $\rmd_\tau^2H=H^{-1}-H$.  The vertical oscillator
therefore has a potential energy function $-\ln H+\half H^2+\cst$,
representing the sum of internal energy and gravitational energy, with
equilibrium at $H=1$.  It can in principle support oscillations of any
energy, but this may involve severe compression because of the slow
divergence of the potential as $H\to0$.  The natural frequency of the
oscillator in the linear regime is $\sqrt{2}\,\Omega_0$ (or
$\sqrt{\gamma+1}\,\Omega_0$ for an arbitrary adiabatic exponent
$\gamma$); more generally, it depends on the amplitude of the
oscillation.

The horizontal and vertical oscillators are coupled in the presence of
a warp.  The $|\psi|$ term in equation~(\ref{duu}) represents the
radial pressure gradient that occurs in a warped disc: in the absence
of a warp, the pressure varies with $z$ to provide hydrostatic
equilibrium, and warping of the disc causes a radial gradient to
appear (Fig.~\ref{f:unbalanced}).  The $|\psi|$ term in
equation~(\ref{dww}) represents the horizontal transport of the
vertical momentum of the warped orbital motion.  Finally, the $|\psi|$
term in equation~(\ref{dhh}) represents a contribution to the velocity
divergence from horizontal motion in a warped disc.  These coupled
horizontal and vertical oscillators featured prominently in the
analysis of weakly nonlinear bending waves in Keplerian discs by
\citet{2006MNRAS.365..977O}.

As we have seen, the viscous terms in a warped disc have a complicated
form.  In the absence of a warp, the equations for laminar viscous
flows can be written
\begin{equation}
  \rmd_\tau U-2V=-\alpha H^{-2}U,
\end{equation}
\begin{equation}
  \rmd_\tau V+(2-q)U=-\alpha H^{-2}V,
\end{equation}
\begin{equation}
  \rmd_\tau W=H^{-1}-H-(\alpha_\rmb+{\textstyle\f{4}{3}}\alpha)H^{-2}W,
\end{equation}
\begin{equation}
  \rmd_\tau H=W,
\end{equation}
showing that the horizontal motion is damped by shear viscosity while
the vertical motion is damped by both shear and bulk viscosity.

\subsection{Numerical solutions}
\label{s:numerical_laminar}

\begin{figure*}
\centerline{\epsfysize7cm\epsfbox{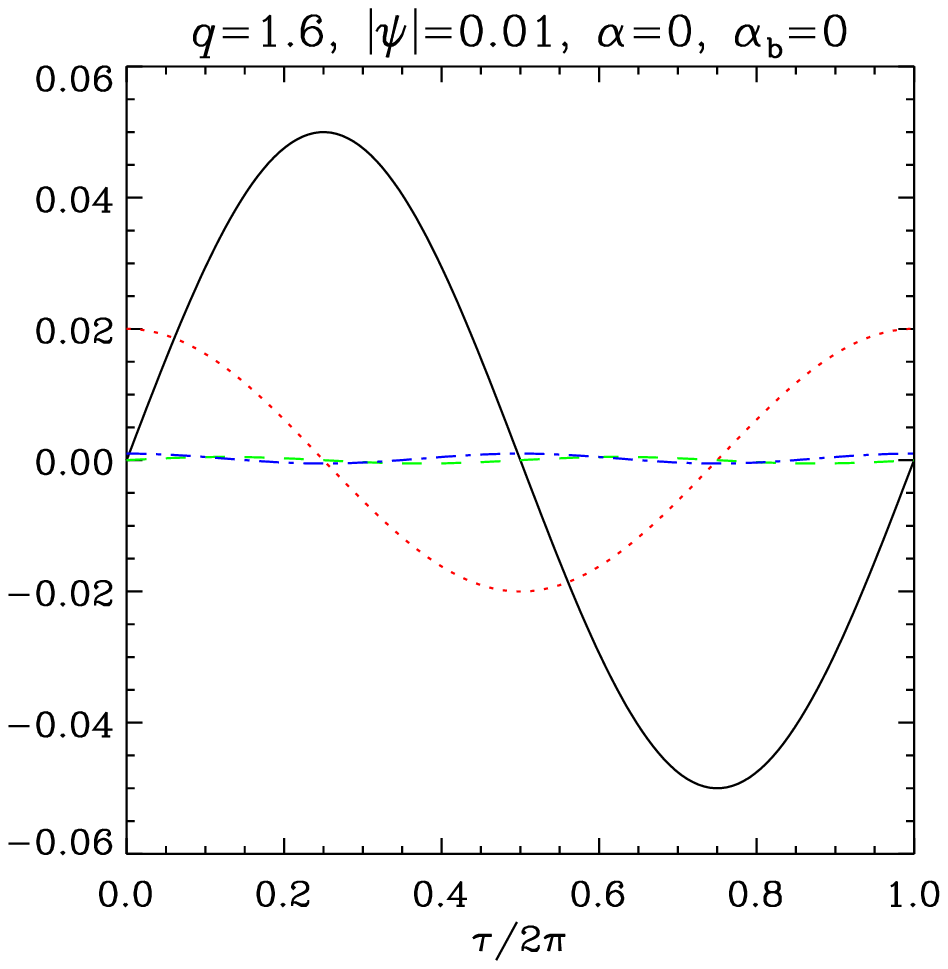}\epsfysize7cm\epsfbox{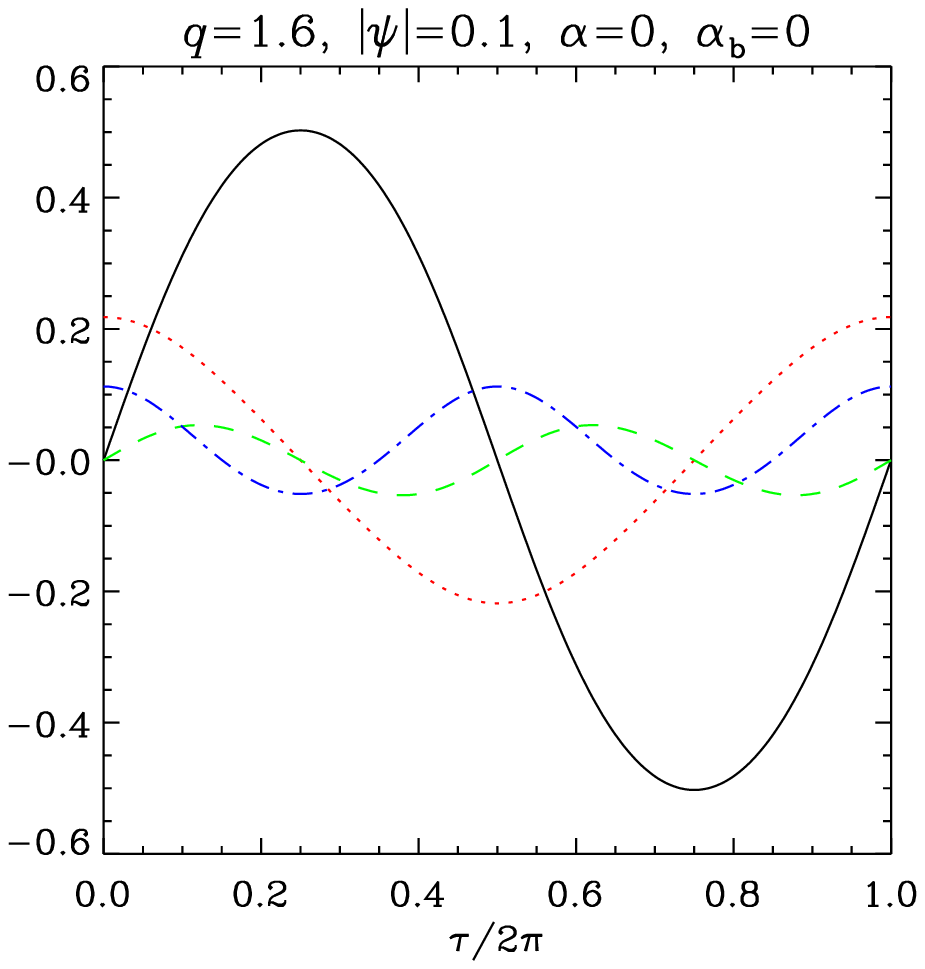}}
\centerline{\epsfysize7cm\epsfbox{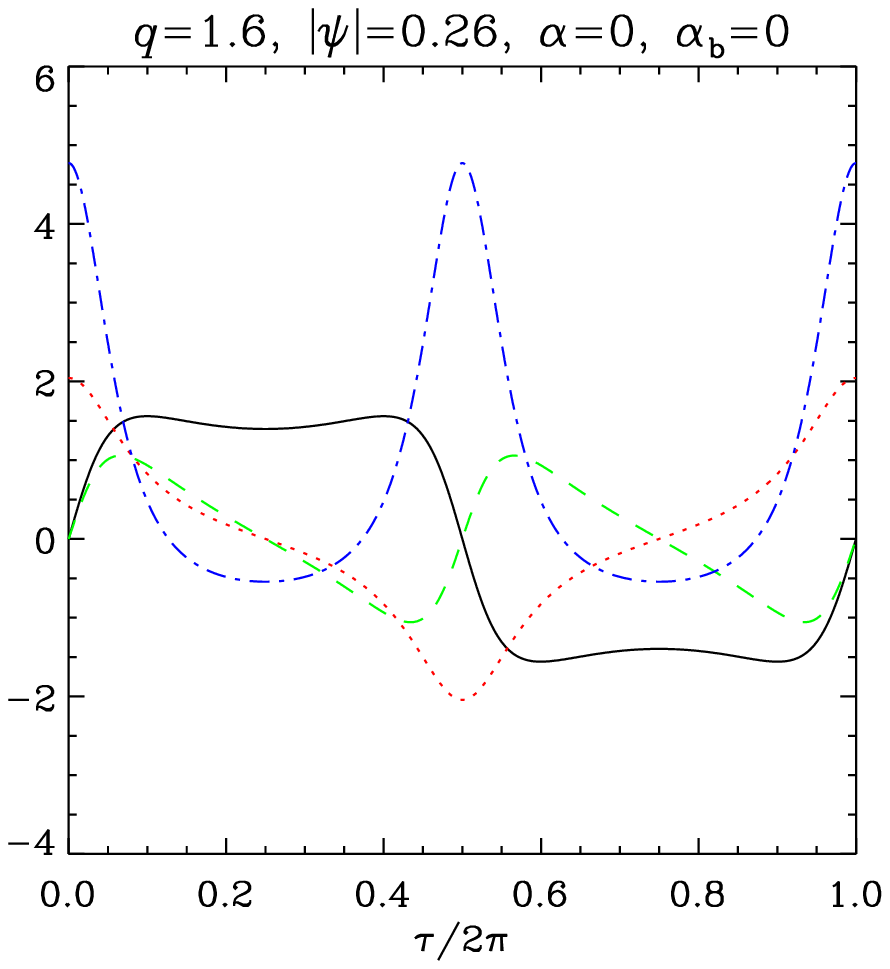}\epsfysize7cm\epsfbox{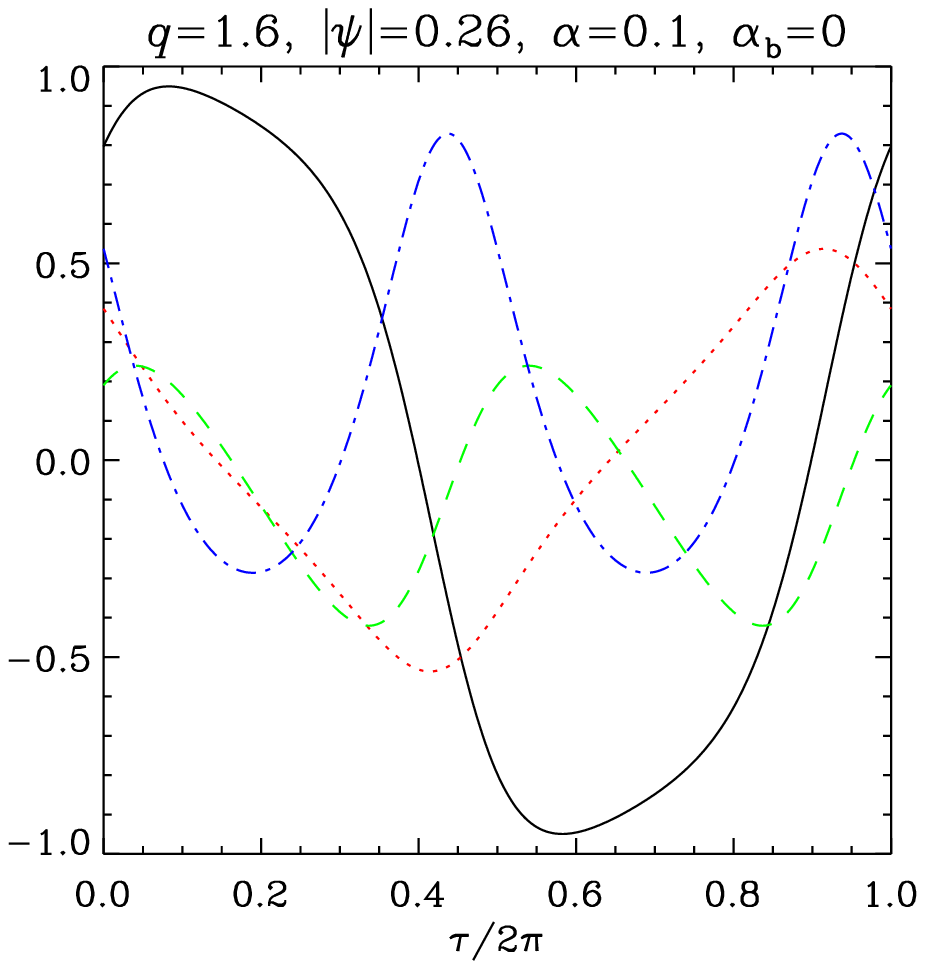}}
\centerline{\epsfysize7cm\epsfbox{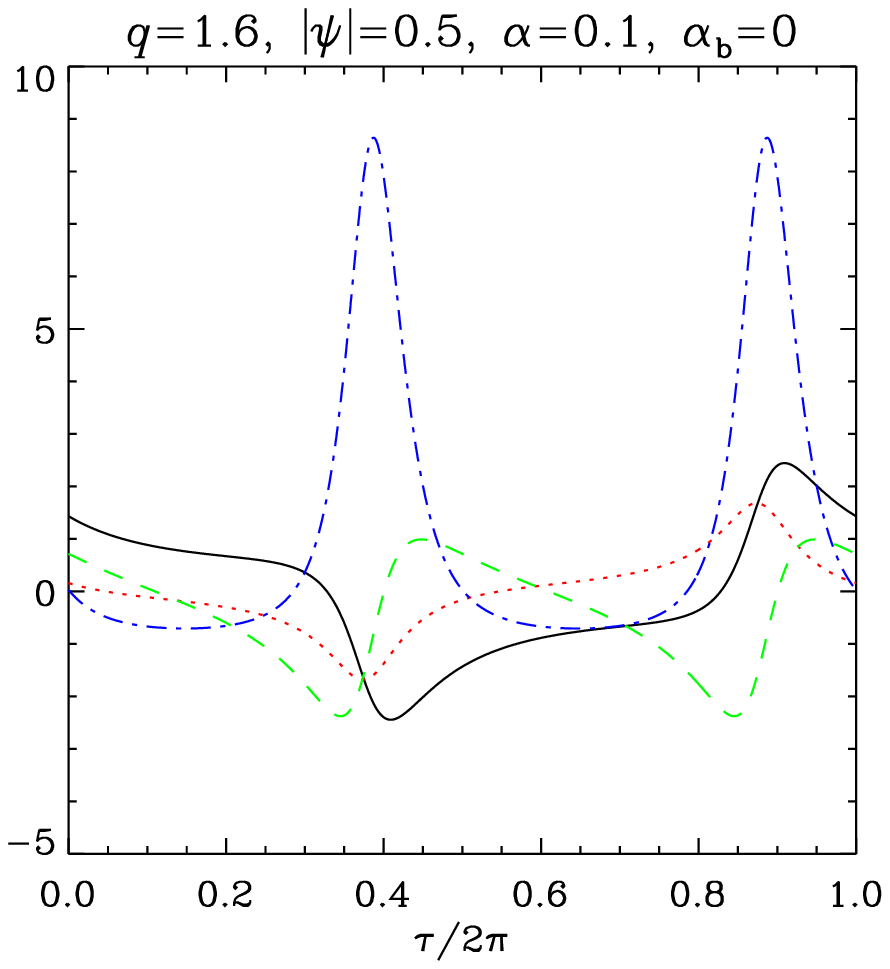}\epsfysize7cm\epsfbox{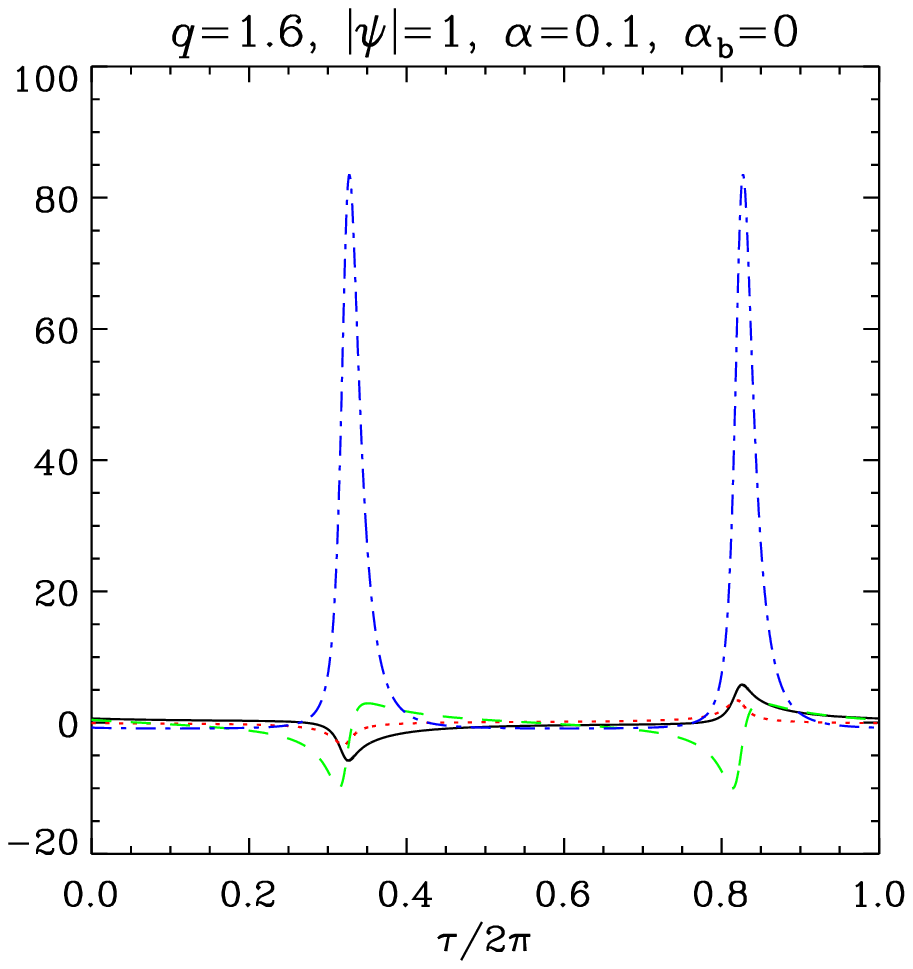}}
\caption{Selection of laminar flows for $q=1.6$.  Solid (black),
  dotted (red), dashed (green) and dot-dashed (blue) lines represent
  $u$, $v$, $w$ and $g-1$, respectively (see
  equations~\ref{vx}--\ref{h}).  The inviscid solutions terminate at
  $|\psi|\approx0.261$ in this case, but larger values of~$|\psi|$ can
  be reached by including viscosity.}
\label{f:laminar_q=1.6}
\end{figure*}

\begin{figure*}
\centerline{\epsfysize7cm\epsfbox{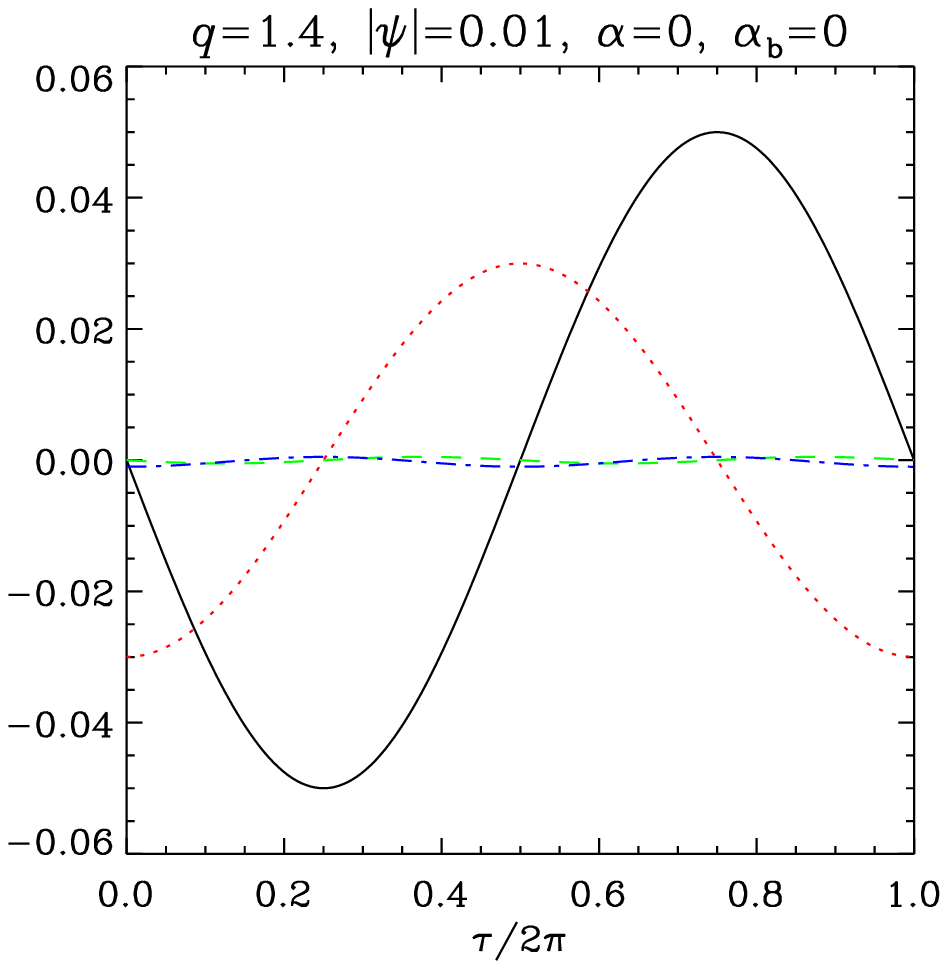}\epsfysize7cm\epsfbox{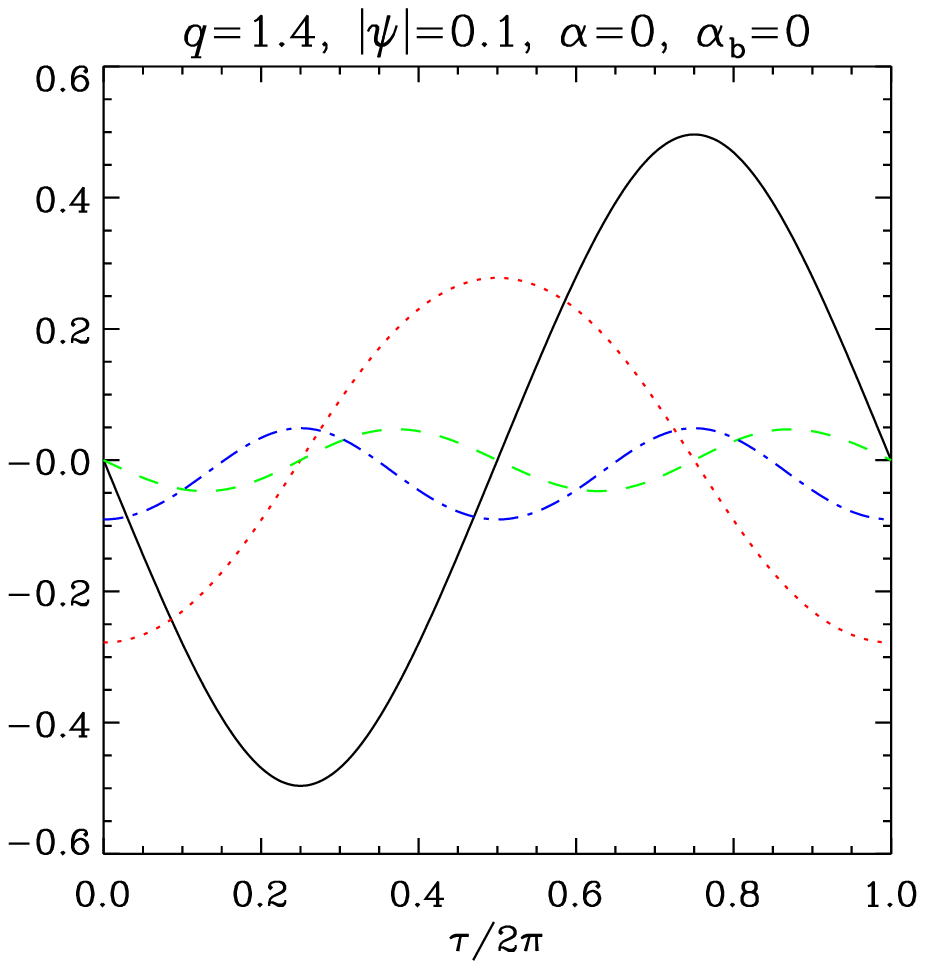}}
\centerline{\epsfysize7cm\epsfbox{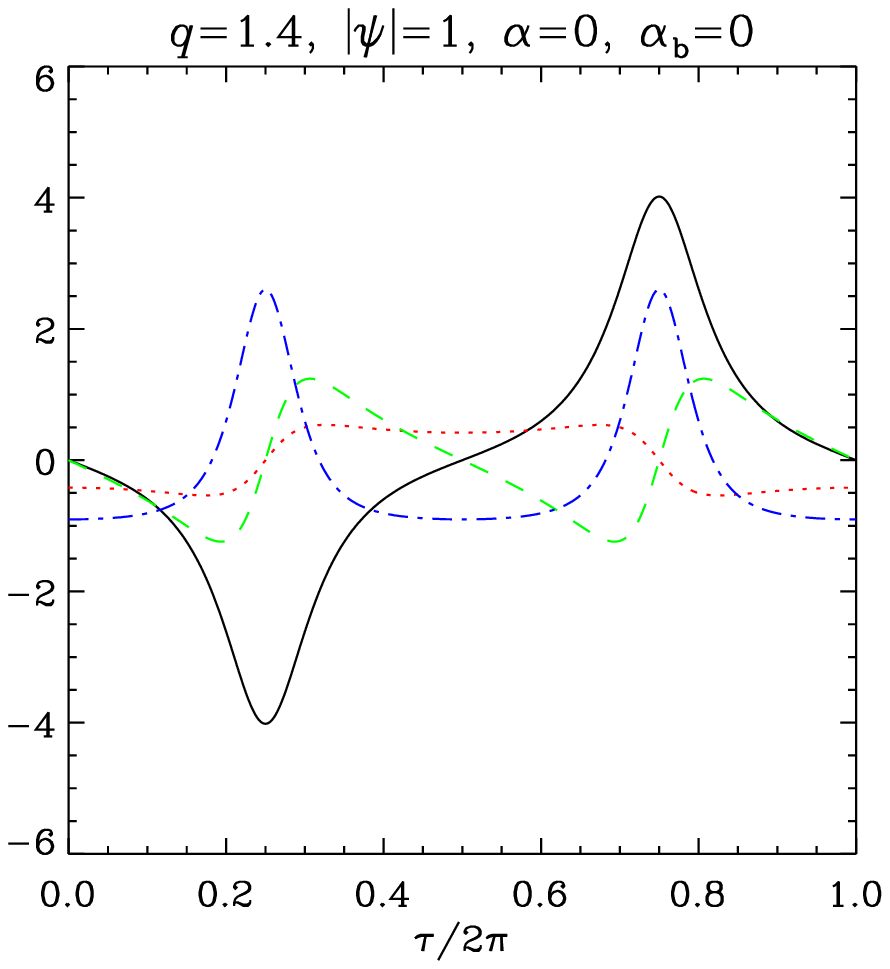}\epsfysize7cm\epsfbox{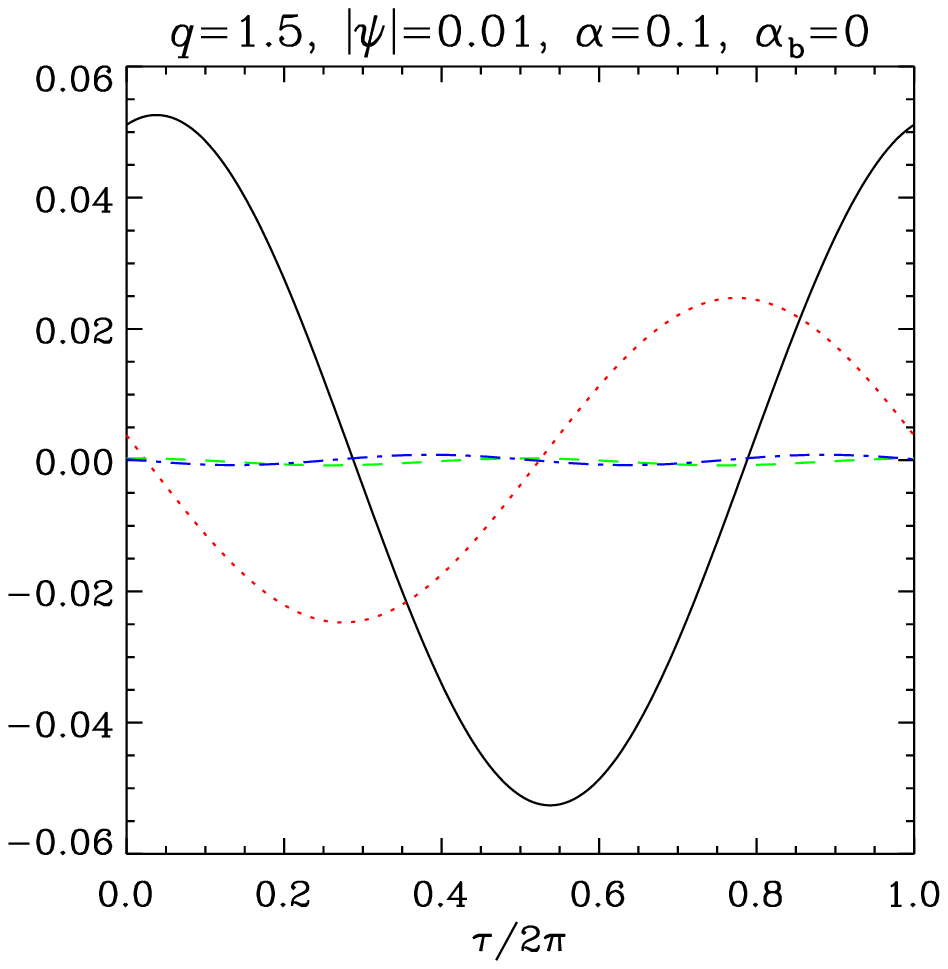}}
\centerline{\epsfysize7cm\epsfbox{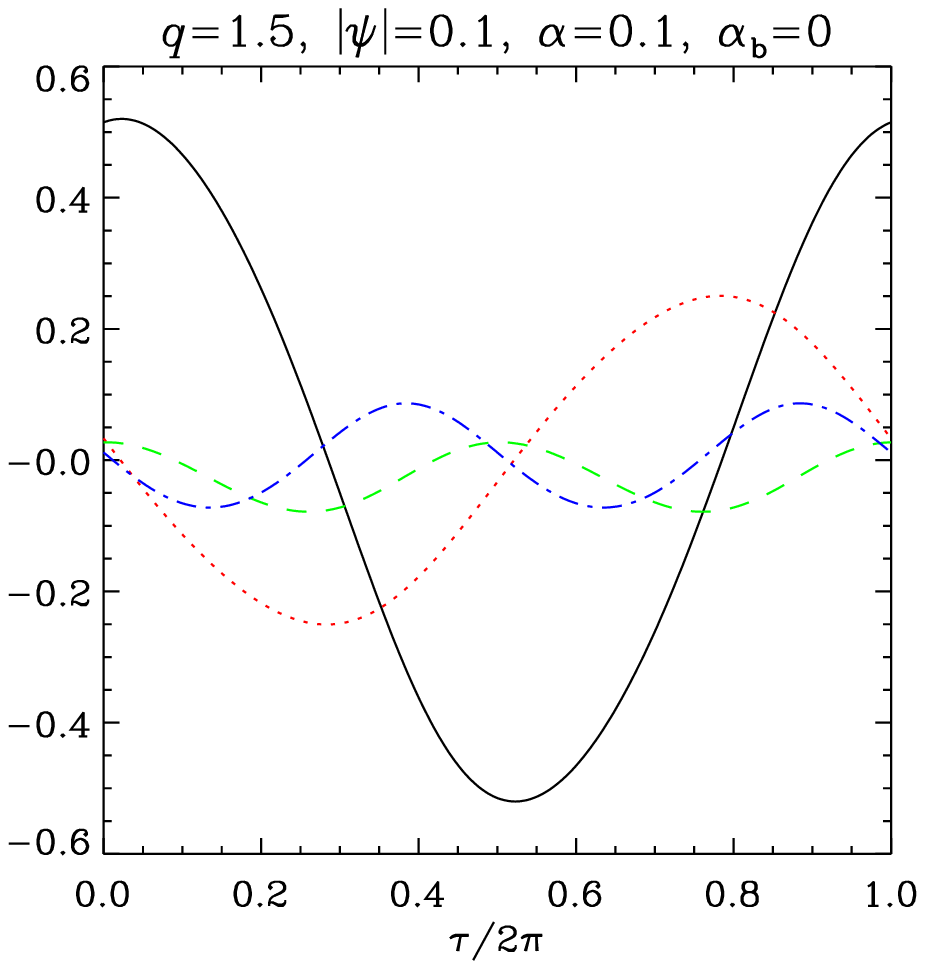}\epsfysize7cm\epsfbox{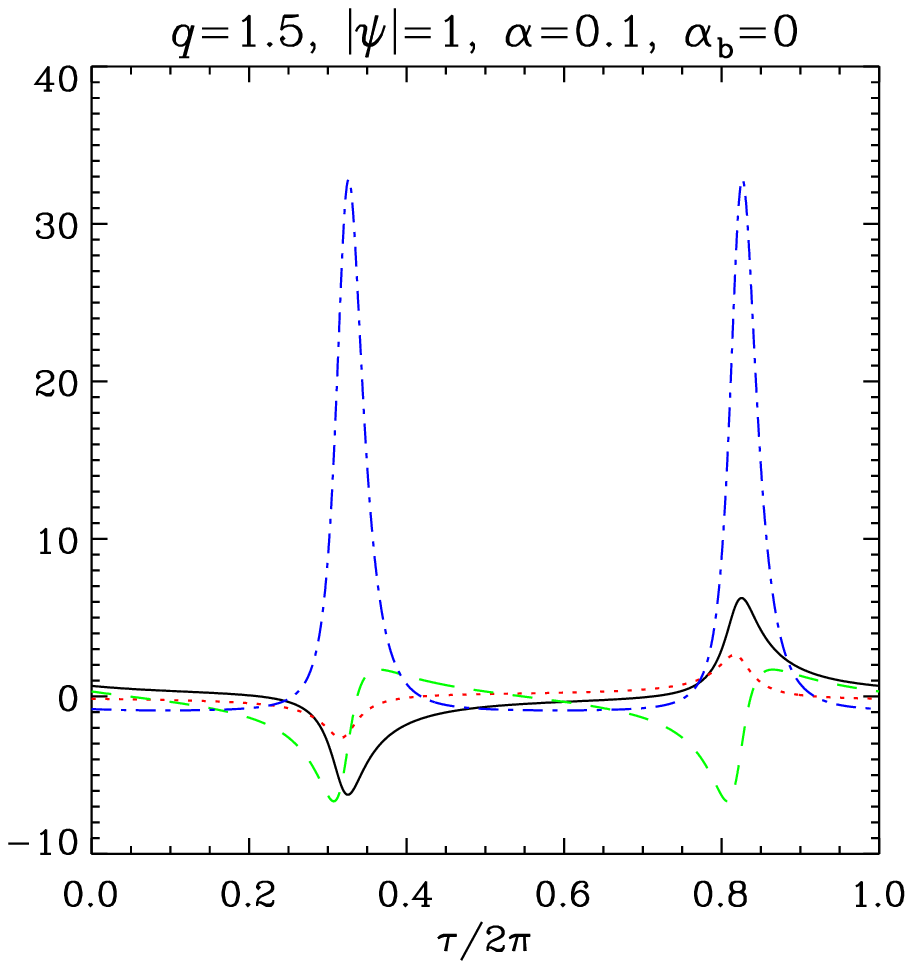}}
\caption{Continuation of Fig.~\ref{f:laminar_q=1.6} for $q=1.4$ and
  $q=1.5$.  The inviscid solutions for $q=1.4$ continue to large
  values of~$|\psi|$.  For $q=1.5$ some viscosity is required.}
\label{f:laminar_q=1.4_q=1.5}
\end{figure*}

Typical numerical solutions for the laminar flows are shown in
Figs~\ref{f:laminar_q=1.6} and~\ref{f:laminar_q=1.4_q=1.5}.  These
were computed by a standard shooting method using a Runge--Kutta
integrator with adaptive stepsize.  We first consider the inviscid
case $\alpha=\alpha_\rmb=0$ for a non-Keplerian disc with $q=1.6$
(Fig.~\ref{f:laminar_q=1.6}, panels 1--3).  Note that the regime
${\textstyle\f{3}{2}}<q<2$, in which the epicyclic frequency is
positive but less than the orbital frequency, corresponds to
conditions close to a black hole, where relativistic effects are
important.  For sufficiently small warp amplitude $|\psi|$, the
horizontal flows $u$ and~$v$ are proportional to $|\psi|$ and have a
sinusoidal dependence on orbital phase, corresponding to the linear
horizontal oscillator being driven at the orbital frequency, which
here is greater than the natural (epicyclic) frequency of the
oscillator (panels 1--2).  The vertical velocity $w$ and the departure
from hydrostatic equilibrium, $g-1$, are proportional to $|\psi|^2$
and therefore smaller.  (Note that $w$ does not include the vertical
velocity associated with the warped orbital motion itself.)

As the warp amplitude is increased further, however, the behaviour of
the solutions changes markedly and the branch of inviscid solutions
terminates; this happens at $|\psi|\approx0.261$ in the case $q=1.6$
(panel~3).  This phenomenon was already noted by
\citet{1999MNRAS.304..557O} and is shown in fig.~2 of that paper.  In
Appendix~\ref{s:existence} we explore mathematically the breakdown of
the inviscid solution and find that this occurs at
$|\psi|^2\approx(q-{\textstyle\f{3}{2}})/1.45$ for
$0<(q-{\textstyle\f{3}{2}})\ll1$.  The physical reason for the
breakdown is a type of nonlinear resonance involving the coupled
horizontal and vertical oscillators.  If a viscosity is included, it
is possible to avoid the breakdown and obtain solutions for larger
values of~$|\psi|$, but they can have quite extreme properties (panels
4--6).

For non-Keplerian discs with $q<{\textstyle\f{3}{2}}$ no such
breakdown occurs and the inviscid solutions can be continued to large
values of~$|\psi|$ (Fig.~\ref{f:laminar_q=1.4_q=1.5}, panels 1--3).
The phases of the horizontal velocities differ by $\pi$ from the case
$q<{\textstyle\f{3}{2}}$ because the orbital frequency is now less
than the epicyclic frequency.

The Keplerian case $q={\textstyle\f{3}{2}}$ is of course of greatest
interest.  Here no $2\pi$-periodic solutions can be obtained without
introducing a viscosity to moderate the resonance resulting from the
coincidence of orbital and epicyclic frequencies.  For sufficiently
small $|\psi|$, the horizontal velocities $u$ and~$v$ are sinusoidal
and proportional to $|\psi|/\alpha$; their phases are intermediate
between those of the inviscid flows in the cases
$q>{\textstyle\f{3}{2}}$ and $q<{\textstyle\f{3}{2}}$ because the
horizontal oscillator is being driven at its natural frequency, rather
than above or below it (panels 4--5).  The solutions can be continued
to large values of~$|\psi|$, but again they can have quite extreme
properties (panel~6).

\subsection{Role of time-dependence of the warp}
\label{s:role}

In the various situations in which warped discs occur, the evolution
of the shape of the disc may take different forms, such as rigid
precession, propagating bending waves, or relaxation towards a
stationary shape.  However, it will usually do so on a time-scale that
is long compared to the orbital time-scale.  This is the justification
of the assumption we made in constructing the local model of a warped
disc, that the geometry of the warp is fixed in a non-rotating frame
of reference.

However, because of the anomalous behaviour of Keplerian warped discs,
it is possible that even a slow time-dependence of the shape of the
disc may have an effect on the local dynamics.  This is because the
horizontal forcing due to the warp no longer occurs at exactly the
epicyclic frequency in a Keplerian disc, and the resonance is slightly
detuned.  A simple way of estimating this effect within the present
framework is to allow the parameter $q$ to be adjusted from its
resonant value of~${\textstyle\f{3}{2}}$.  If, for example, the warp
precesses slowly at a rate $\Omega_\rmp$ ($<0$ for retrograde
precession), which is determined by global considerations, then in the
fluid frame the geometry oscillates at frequency $\Omega-\Omega_\rmp$.
If the disc is Keplerian, then the required detuning of~$\Omega_\rmp$
between the horizontal forcing frequency and the epicyclic frequency
can be achieved by adjusting $q$ such that
$[\sqrt{2(2-q)}-1]\Omega=\Omega_\rmp$.  This adjustment is very small
if, as is usually the case, $|\Omega_\rmp|\ll\Omega$, but it is
possible that the detuning it provides may limit the amplitude of the
laminar flows if $\alpha$ is also very small.

\subsection{Torques}
\label{s:torques}

In Section~\ref{s:fluxes} we showed how to compute the internal torque
in the local model.  For the laminar viscous flows described in
Section~\ref{s:separation}, the dimensionless torque coefficients
evaluate to
\begin{equation}
  Q_1=\langle-g^{-1}uv+\alpha(-q+|\psi|\cos\tau\,v)\rangle_\tau
\end{equation}
and
\begin{equation}
  Q_4|\psi|=\langle\rme^{\rmi\tau}[g^{-1}u(1+\rmi w)-\rmi\alpha(|\psi|\sin\tau+|\psi|\cos\tau\,w+u)]\rangle_\tau.
\end{equation}
These are the direct analogues of equations~(112) and~(120) in
\citet{1999MNRAS.304..557O}, but are expressed in a different
notation, and are slightly simplified because the disc is isothermal.
The case of adiabatic flow is treated in Appendix~\ref{s:adiabatic},
where we obtain expressions that are exactly equivalent to those of
\citet{1999MNRAS.304..557O} and we explain the notational
correspondence.

Note that, if the laminar flow is neglected by setting $u=v=w=0$, we
obtain the simple but incorrect results $Q_1=-\alpha q$ and
$Q_4=\half\alpha$.  These torque coefficients represent only the
viscous stresses associated with the warped orbital motion, and would
lead to a diffusion of the warp on (twice) the viscous timescale as
found erroneously in early theoretical work on warped discs.  In
practice the torque (mainly from the radial advective transport of
horizontal angular momentum) associated with the laminar flows exceeds
the viscous torque, and leads to a more rapid evolution of the warp.

\begin{figure}
\centerline{\epsfysize8cm\epsfbox{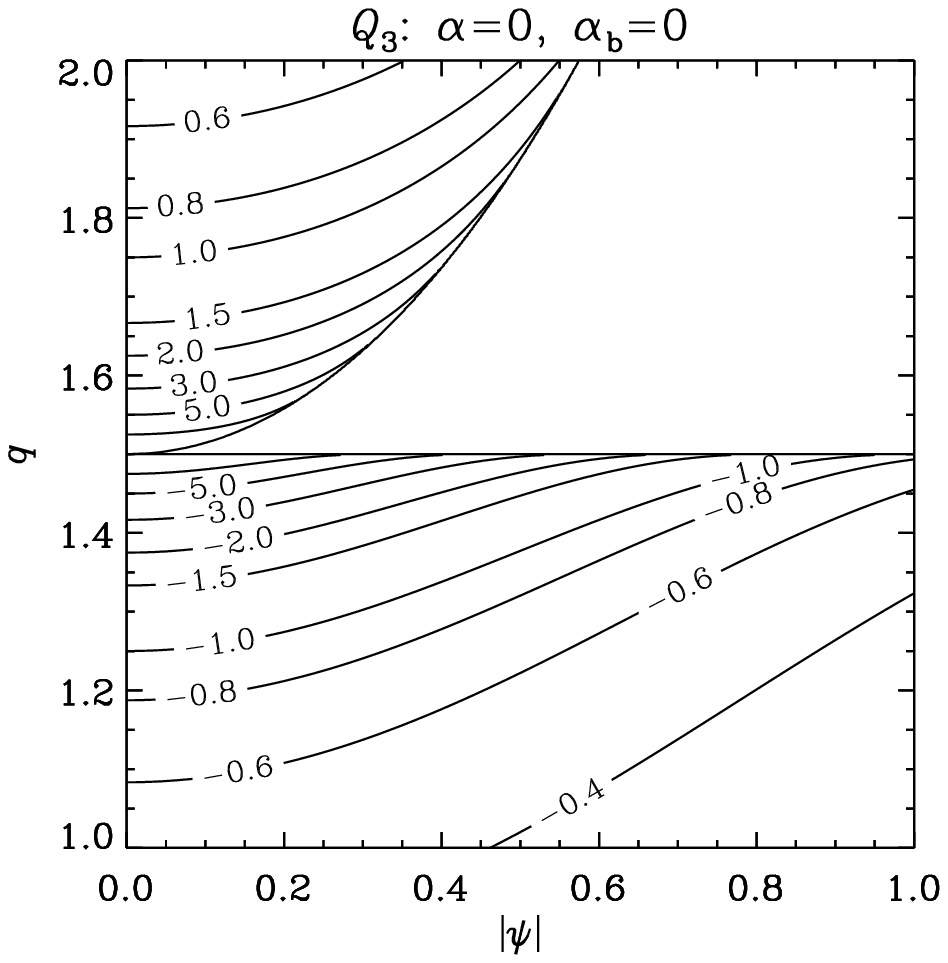}}
\caption{Dependence of the torque coefficient $Q_3$ of laminar flows
  on the warp amplitude and dimensionless shear rate for an inviscid
  non-Keplerian disc.  In the blank sector (top right) no solutions
  are found.  The short unlabelled contours have values $\pm10$.}
\label{f:q3inviscid}
\end{figure}

For an inviscid disc, which must be non-Keplerian for laminar
solutions to exist, only $Q_3$ is non-zero and it determines the
(dispersive) propagation of bending waves according to the equation
\begin{equation}
  \Sigma r^3\Omega\f{\p\bl}{\p t}=\f{\p}{\p r}\left(Q_3\Sigma c_\rms^2r^3\,\bl\times\f{\p\bl}{\p r}\right),
\end{equation}
which follows from equation~(\ref{shape}) with $\calF=0$ in this case.
The variation of~$Q_3$ with $|\psi|$ and~$q$ is shown in
Fig.~\ref{f:q3inviscid}, which is equivalent to fig.~2 of
\citet{1999MNRAS.304..557O} except that it is for an isothermal disc.
As noted in Section~\ref{s:numerical_laminar}, the inviscid solutions
break down for ${\textstyle{\f{3}{2}}}<q<2$ when $|\psi|$ exceeds a
critical value, owing to a type of nonlinear resonance.  The
analytical result for sufficiently small $|\psi|^2/(2q-3)$ is
\begin{equation}
  Q_3=\f{1}{2(2q-3)}\left[1+\f{5}{4}\f{|\psi|^2}{(2q-3)}+O\left(\f{|\psi|^4}{(2q-3)^2}\right)\right].
\end{equation}

\begin{figure*}
\centerline{\epsfysize7.5cm\epsfbox{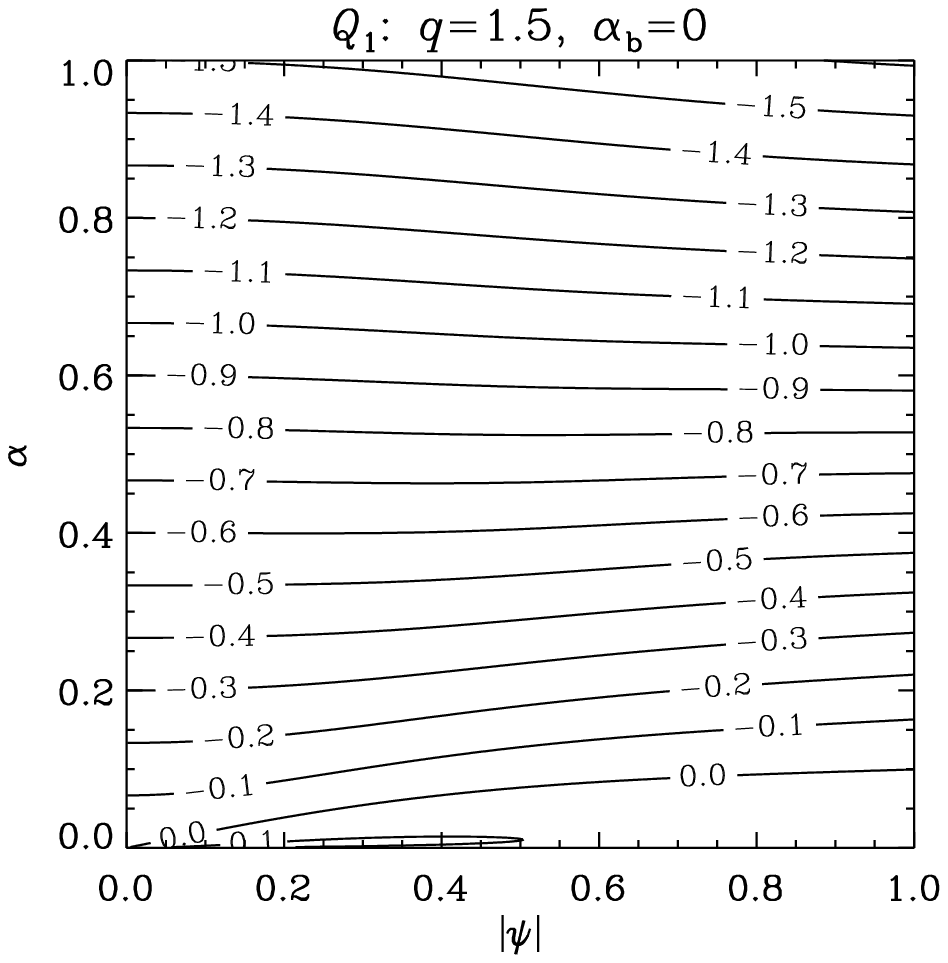}\qquad\epsfysize7.5cm\epsfbox{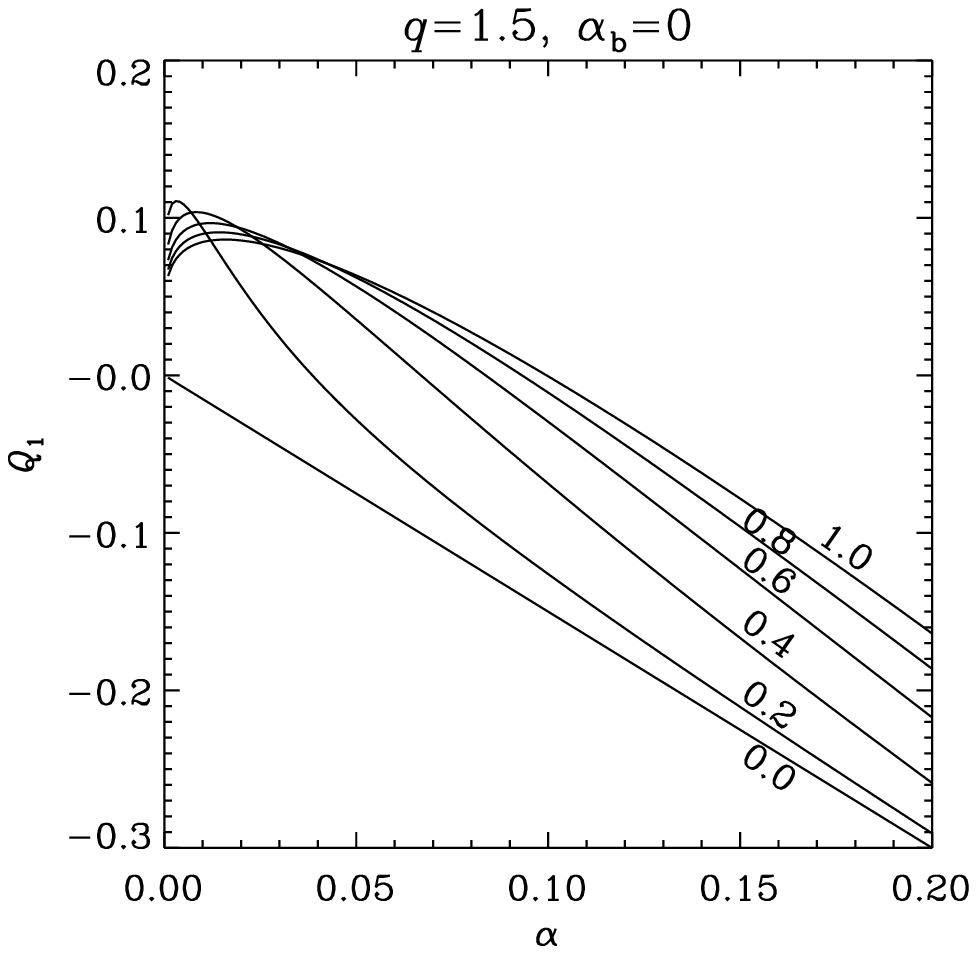}}
\centerline{\epsfysize7.5cm\epsfbox{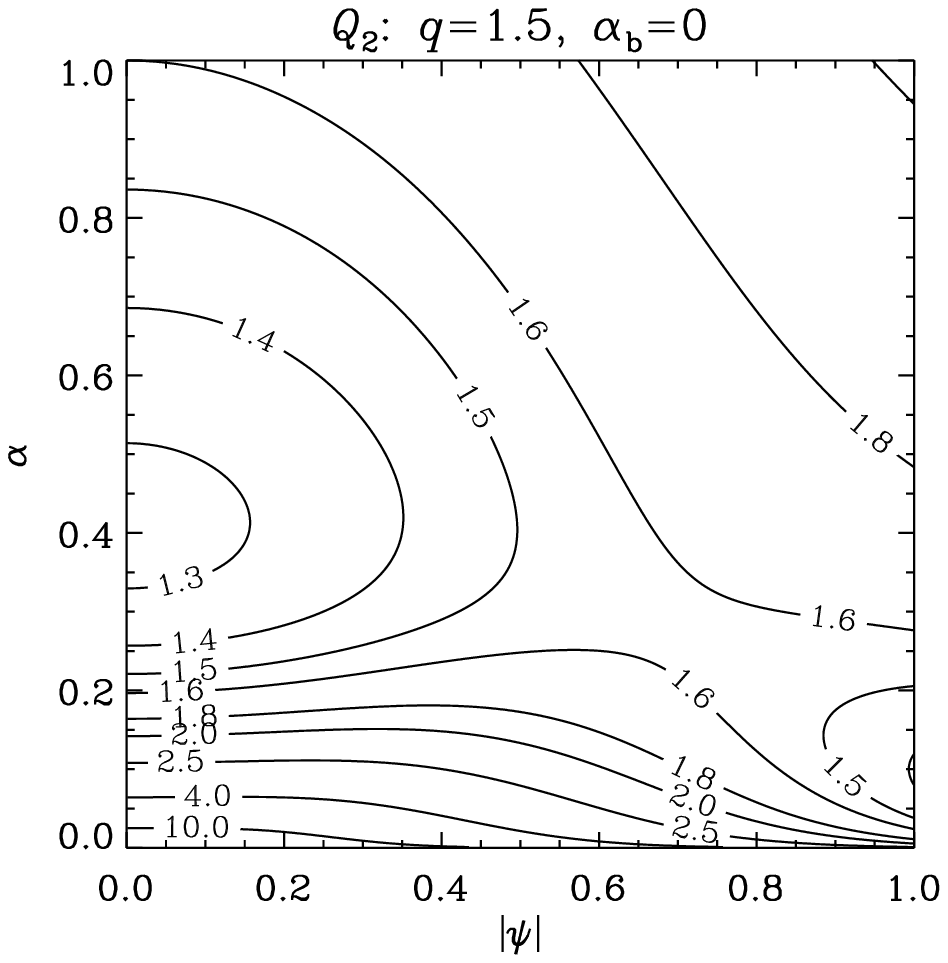}\qquad\epsfysize7.5cm\epsfbox{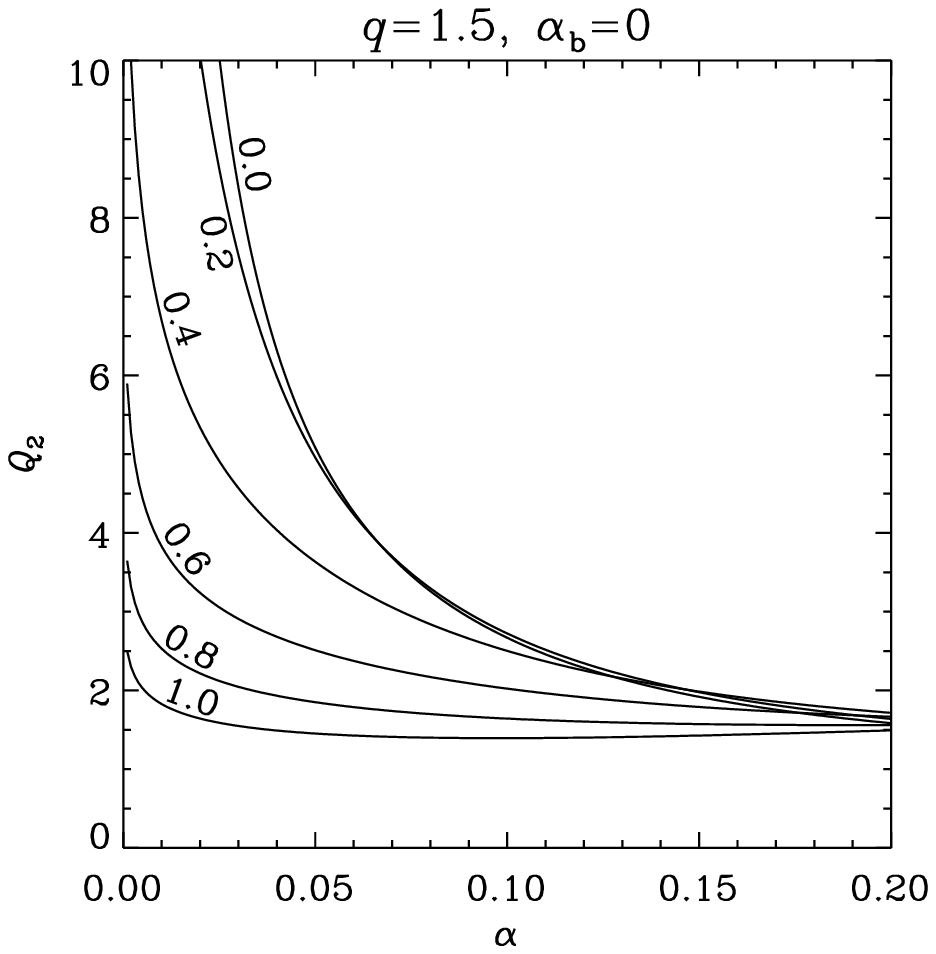}}
\centerline{\epsfysize7.5cm\epsfbox{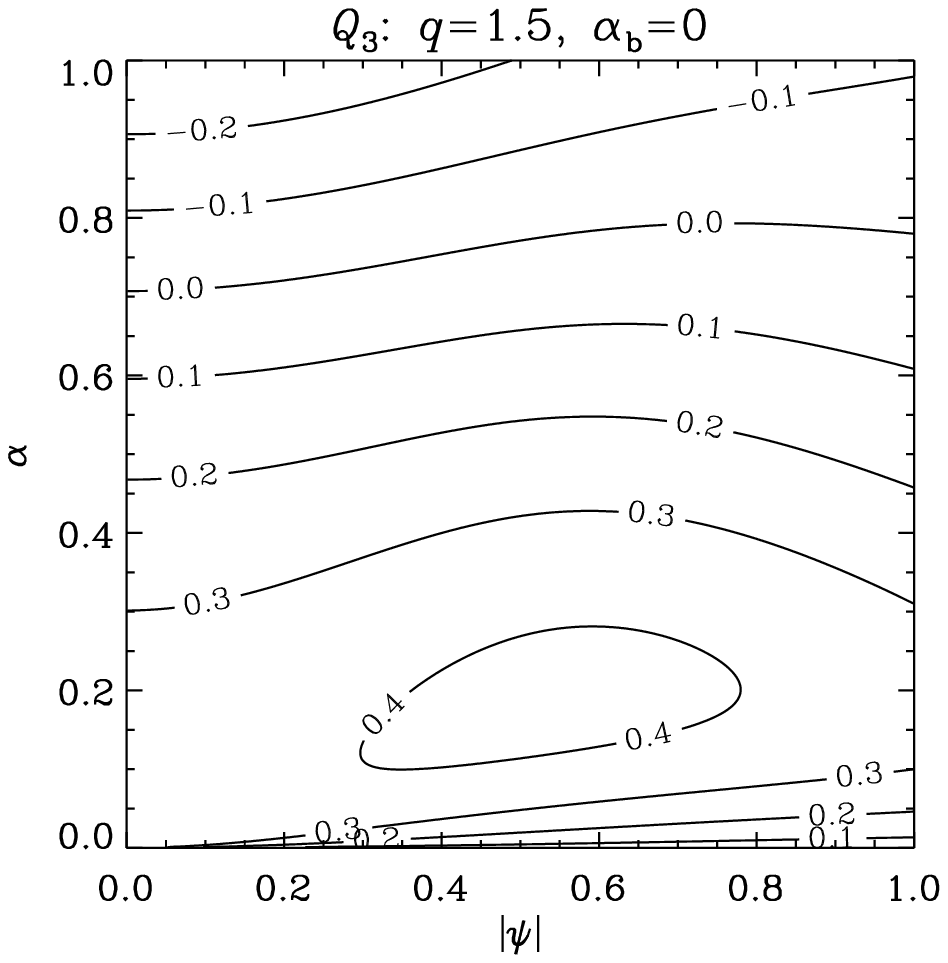}\qquad\epsfysize7.5cm\epsfbox{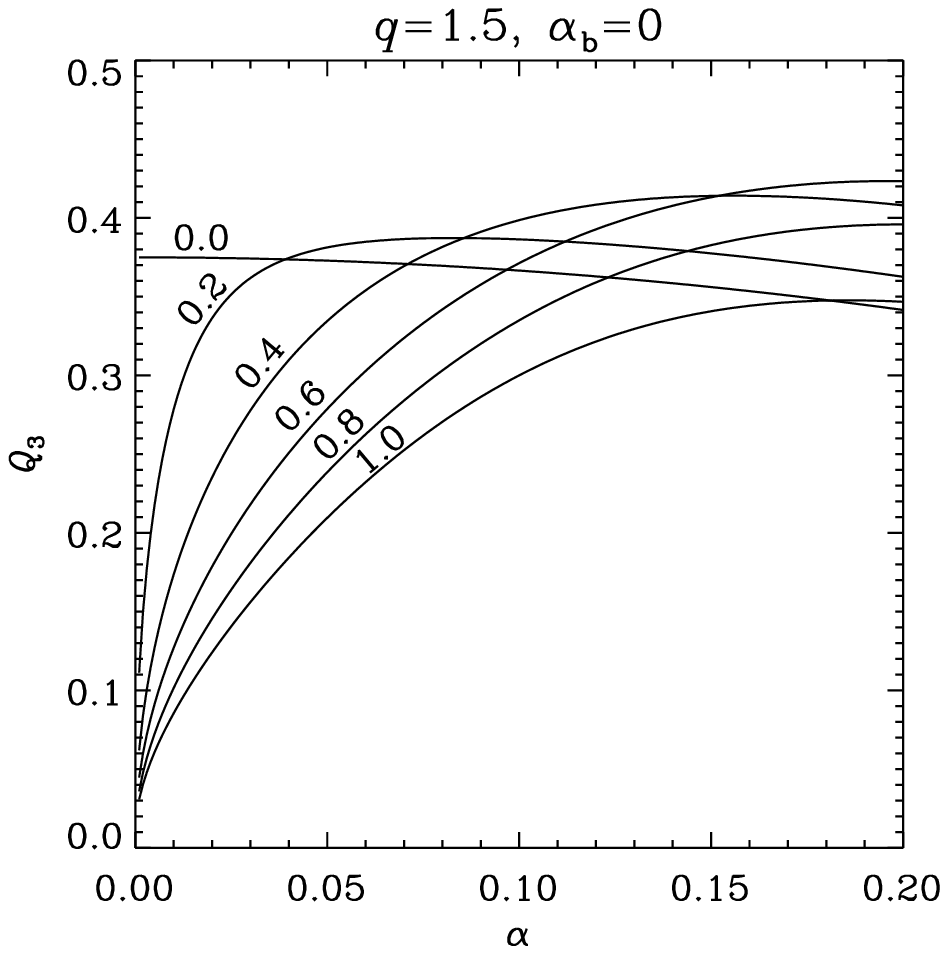}}
\caption{Dependence of the torque coefficients $Q_1$ (top), $Q_2$
  (middle) and $Q_3$ (bottom) of laminar flows on the warp amplitude
  and viscosity for a Keplerian disc without bulk viscosity.  The
  right-hand panels show the torque coefficients as functions of
  viscosity for $0<\alpha<0.2$ for various values of the warp
  amplitude as labelled on each curve.}
\label{f:q1q2q3kep}
\end{figure*}

In Fig.~\ref{f:q1q2q3kep} we show the variation of all three
coefficients with $|\psi|$ and~$\alpha$ for a Keplerian disc without
bulk viscosity.  These results are equivalent to figs 3--5 of
\citet{1999MNRAS.304..557O} except that they are for an isothermal
disc.  The analytical results for sufficiently small $|\psi|^2$ in
this case are
\begin{equation}
  Q_1=-\f{3\alpha}{2}+\f{(1-8\alpha^2+3\alpha^4)}{4\alpha(4+\alpha^2)}|\psi|^2+O(|\psi|^4)
\end{equation}
and
\begin{equation}
  Q_4=\f{1+2\rmi\alpha+6\alpha^2}{2\alpha(2+\rmi\alpha)}+O(|\psi|^2),
\end{equation}
which can be decomposed into
\begin{equation}
  Q_2=\f{1+7\alpha^2}{\alpha(4+\alpha^2)}+O(|\psi|^2),
\end{equation}
\begin{equation}
  Q_3=\f{3-6\alpha^2}{2(4+\alpha^2)}+O(|\psi|^2).
\end{equation}
The figures show that the variation with amplitude is mild except when
the viscosity is small, as highlighted in the right-hand panels.  In
particular, the coefficient $Q_1$ reverses sign for sufficiently large
$|\psi|$ and small $\alpha$.  This angular-momentum flux reversal
occurs when the Reynolds stress associated with the correlation of~$u$
and~$v$ exceeds the usual viscous stress.  If it occurs, it can be
expected to cause an antidiffusion of the mass distribution of the
disc, leading to the formation of disjoint
rings.\footnote{Angular-momentum flux reversal occurs in planetary
  rings that are strongly perturbed by nonlinear density waves, and is
  thought to be responsible for the formation of sharp edges
  \citep{1989Icar...80..344B}.}  The warp diffusion coefficient $Q_2$
diverges at small $\alpha$ and small $|\psi|$ where the resonance is
located, but this effect is weakened by nonlinearity at larger
amplitude.  As shown in Paper~II, however, the laminar solutions are
generally unstable at small $\alpha$ and large $|\psi|$, so the torque
coefficients cannot be trusted in this region.

\section{Conclusion}

This paper offers several advances in the theory of warped discs.
First, we have constructed a local model that generalizes the shearing
sheet of \citet{1965MNRAS.130..125G}.  The warped shearing sheet is
horizontally homogeneous and admits periodic boundary conditions,
although its geometry oscillates at the orbital frequency.  This leads
to a computational model in the form of a warped shearing box, which
can be used for linear and nonlinear studies of fluid dynamics,
magnetohydrodynamics, etc.\ in warped discs, especially to investigate
processes such as instability and turbulence that cannot be resolved
in global numerical simulations.  It is related to the standard
shearing box by a relatively simple coordinate transformation
described in Section~\ref{s:coordinates}.  Second, we have shown how
to use the local model to calculate all three components of the torque
that governs the large-scale evolution of the shape and mass
distribution of the disc.  Third, we have provided an independent and
simpler route to the nonlinear theory of warped discs derived by
\citet{1999MNRAS.304..557O}, showing how the solutions obtained in
that paper correspond to the simplest laminar flows that can occur in
the local model.  In Paper~II we use the local model to analyse the
linear hydrodynamic stability of these laminar flows.  The widespread
instability that we find there requires further investigation in
future work.

\section*{acknowledgments}

This research was supported by STFC.  We are grateful for the
referee's suggestions.

\newpage

\appendix

\section{Adiabatic flow}
\label{s:adiabatic}

The basic equations governing the adiabatic flow of an inviscid fluid
in the local model are
\begin{equation}
  \rmD u_x-2\Omega_0u_y=2q\Omega_0^2x-\f{1}{\rho}\p_xp,
\end{equation}
\begin{equation}
  \rmD u_y+2\Omega_0u_x=-\f{1}{\rho}\p_yp,
\end{equation}
\begin{equation}
  \rmD u_z=-\Omega_0^2z-\f{1}{\rho}\p_zp,
\end{equation}
\begin{equation}
  \rmD\rho=-\rho(\p_xu_x+\p_yu_y+\p_zu_z),
\end{equation}
\begin{equation}
  \rmD p=-\gamma p(\p_xu_x+\p_yu_y+\p_zu_z),
\end{equation}
where
\begin{equation}
  \rmD=\p_t+u_x\p_x+u_y\p_y+u_z\p_z
\end{equation}
is the Lagrangian derivative and $\gamma$ is the adiabatic exponent,
which we assume to be constant.  By introducing the specific internal
energy $e$, the specific enthalpy $h$, the specific entropy $s$ and
the temperature $T$, which satisfy the relations
\begin{equation}
  h=e+\f{p}{\rho},
\end{equation}
\begin{equation}
  \rmd h=T\,\rmd s+\f{\rmd p}{\rho},
\end{equation}
the basic equations can be rewritten in the form
\begin{equation}
  \rmD u_x-2\Omega_0u_y=2q\Omega_0^2x-\p_xh+T\p_xs,
\end{equation}
\begin{equation}
  \rmD u_y+2\Omega_0u_x=-\p_yh+T\p_ys,
\end{equation}
\begin{equation}
  \rmD u_z=-\Omega_0^2z-\p_zh+T\p_zs,
\end{equation}
\begin{equation}
  \rmD h=-(\gamma-1)h(\p_xu_x+\p_yu_y+\p_zu_z),
\end{equation}
\begin{equation}
  \rmD s=0,
\end{equation}
assuming an ideal gas for which $h=\gamma e$.  In warped shearing
coordinates, these equations take the form
\begin{eqnarray}
  \lefteqn{\rmD v_x-2\Omega_0v_y=-(\p_x'+q\tau\,\p_y'+|\psi|\cos\tau\,\p_z')h}&\nonumber\\
  &&+T(\p_x'+q\tau\,\p_y'+|\psi|\cos\tau\,\p_z')s,
\end{eqnarray}
\begin{equation}
  \rmD v_y+(2-q)\Omega_0v_x=-\p_y'h+T\p_y's,
\end{equation}
\begin{equation}
  \rmD v_z+|\psi|\Omega_0\sin\tau\,v_x=-\Omega_0^2z'-\p_z'h+T\p_z's,
\end{equation}
\begin{eqnarray}
  \lefteqn{\rmD h=-(\gamma-1)h[(\p_x'+q\tau\,\p_y'+|\psi|\cos\tau\,\p_z')v_x+\p_y'v_y+\p_z'v_z],}&\nonumber\\
\end{eqnarray}
\begin{equation}
  \rmD s=0,
\end{equation}
with
\begin{equation}
  \rmD=\p_t'+v_x\p_x'+(v_y+q\tau v_x)\p_y'+(v_z+|\psi|\cos\tau\,v_x)\p_z',
\end{equation}
where $\bv$ is the relative velocity.  Laminar flows independent
of~$x'$ and~$y'$ satisfy
\begin{equation}
  \rmD v_x-2\Omega_0v_y=-|\psi|\cos\tau\,\p_z'h+T|\psi|\cos\tau\,\p_z's,
\end{equation}
\begin{equation}
  \rmD v_y+(2-q)\Omega_0v_x=0,
\end{equation}
\begin{equation}
  \rmD v_z+|\psi|\Omega_0\sin\tau\,v_x=-\Omega_0^2z'-\p_z'h+T\p_z's,
\end{equation}
\begin{equation}
  \rmD h=-(\gamma-1)h(|\psi|\cos\tau\,\p_z'v_x+\p_z'v_z),
\end{equation}
\begin{equation}
  \rmD s=0,
\end{equation}
with
\begin{equation}
  \rmD=\p_t'+(v_z+|\psi|\cos\tau\,v_x)\p_z'.
\end{equation}
A nonlinear separation of variables is then possible, with
\begin{equation}
  v_x(z',t')=u(\tau)\Omega_0z',
\end{equation}
\begin{equation}
  v_y(z',t')=v(\tau)\Omega_0z',
\end{equation}
\begin{equation}
  v_z(z',t')=w(\tau)\Omega_0z',
\end{equation}
\begin{equation}
  h(z',t')=\Omega_0^2[f(\tau)-\half g(\tau)z'^2],
\end{equation}
\begin{equation}
  s(z',t')=s(\tau),
\end{equation}
where $u$, $v$, $w$ and~$g$ are all dimensionless.  Thus
\begin{equation}
  \rmd_\tau u+(w+|\psi|\cos\tau\,u)u-2v=|\psi|\cos\tau\,g,
\end{equation}
\begin{equation}
  \rmd_\tau v+(w+|\psi|\cos\tau\,u)v+(2-q)u=0,
\end{equation}
\begin{equation}
  \rmd_\tau w+(w+|\psi|\cos\tau\,u)w+|\psi|\sin\tau\,u=g-1,
\end{equation}
\begin{equation}
  \rmd_\tau f=-(\gamma-1)(w+|\psi|\cos\tau\,u)f,
\end{equation}
\begin{equation}
  \rmd_\tau g=-(\gamma+1)(w+|\psi|\cos\tau\,u)g,
\end{equation}
\begin{equation}
  \rmd_\tau s=0.
\end{equation}
When viscosity is included in the equation of motion, but viscous
heating is neglected, these equations become
\begin{eqnarray}
  \lefteqn{\rmd_\tau u+(w+|\psi|\cos\tau\,u)u-2v=|\psi|\cos\tau\,g}&\nonumber\\
  &&-(\alpha_\rmb+\third\alpha)|\psi|\cos\tau\,g(w+|\psi|\cos\tau\,u)\nonumber\\
  &&-\alpha g[|\psi|\sin\tau+(1+|\psi|^2\cos^2\tau)u],
\end{eqnarray}
\begin{eqnarray}
  \lefteqn{\rmd_\tau v+(w+|\psi|\cos\tau\,u)v+(2-q)u}&\nonumber\\
  &&=-\alpha g[-q|\psi|\cos\tau+(1+|\psi|^2\cos^2\tau)v],
\end{eqnarray}
\begin{eqnarray}
  \lefteqn{\rmd_\tau w+(w+|\psi|\cos\tau\,u)w+|\psi|\sin\tau\,u=g-1}&\nonumber\\
  &&-(\alpha_\rmb+\third\alpha)g(w+|\psi|\cos\tau\,u)\nonumber\\
  &&-\alpha g[|\psi|^2\sin\tau\cos\tau+(1+|\psi|^2\cos^2\tau)w],
\end{eqnarray}
\begin{equation}
  \rmd_\tau f=-(\gamma-1)(w+|\psi|\cos\tau\,u)f,
\end{equation}
\begin{equation}
  \rmd_\tau g=-(\gamma+1)(w+|\psi|\cos\tau\,u)g,
\end{equation}
\begin{equation}
  \rmd_\tau s=0.
\end{equation}
These are exactly equivalent to equations (105)--(109) of
\citet{1999MNRAS.304..557O}.  The mapping from the earlier notation to
the current notation is as follows: $f_1\mapsto f$, $f_2\mapsto g$,
$f_3\mapsto u$, $f_4\mapsto-(w+|\psi|\cos\tau\,u)$, $f_5\mapsto v$,
$\Gamma\mapsto\gamma$, $\hat\kappa^2\mapsto2(2-q)$, $\phi\mapsto\tau$.
The equations for $f$ and~$g$ are related in such a way that the
surface density of the disc, which is proportional to
$f^{(\gamma+1)/2(\gamma-1)}g^{-1/2}$, is independent of~$\tau$.

The dimensionless torque coefficients for a non-isothermal disc must
be defined differently from equation~(\ref{torque}), because $c_\rms$
is no longer constant.  As in \citet{1999MNRAS.304..557O}, we define
instead
\begin{equation}
  \bcalG=-2\pi\calI r^2\Omega^2\left(Q_1\,\bl+Q_2r\f{\p\bl}{\p r}+Q_3r\,\bl\times\f{\p\bl}{\p r}\right),
\end{equation}
where
\begin{equation}
  \calI=\bigg\langle\int\rho z'^2\,\rmd z'\bigg\rangle_{\!\!\rmh,\tau}
\end{equation}
is the orbitally averaged second vertical moment of the density.
Prior to averaging, its variation with orbital phase is proportional
to $\propto fg^{-1}$, and we define $f_6=fg^{-1}/\langle
fg^{-1}\rangle_\tau$ as in the earlier notation.  Then the
dimensionless torque coefficients for these laminar flows are given by
\begin{equation}
  Q_1=\langle f_6[-uv+\alpha g(-q+|\psi|\cos\tau\,v)]\rangle_\tau,
\end{equation}
\begin{eqnarray}
  \lefteqn{Q_4|\psi|=\langle\rme^{\rmi\tau}f_6[u(1+\rmi w)-\rmi\alpha g(|\psi|\sin\tau+|\psi|\cos\tau\,w+u)]\rangle_\tau,}&\nonumber\\
\end{eqnarray}
which agree with equations~(112) and~(120) in
\citet{1999MNRAS.304..557O}.

\section{Existence of inviscid laminar flow solutions}
\label{s:existence}

Equations (\ref{duu})--(\ref{dhh}) for inviscid laminar flows are
\begin{equation}
  \rmd_\tau U-2V=|\psi|\cos\tau\,H^{-1},
\end{equation}
\begin{equation}
  \rmd_\tau V+(2-q)U=0,
\end{equation}
\begin{equation}
  \rmd_\tau W+|\psi|\sin\tau\,U=H^{-1}-H,
\end{equation}
\begin{equation}
  \rmd_\tau H=W+|\psi|\cos\tau\,U.
\end{equation}
Note that $U$ and~$W$ are expected to be odd in $\tau$ while $V$ and
$H$ are even.

The regular expansion of the solution for small $|\psi|$ is
\begin{equation}
  U=|\psi|U_1+O(|\psi|^3),
\end{equation}
\begin{equation}
  V=|\psi|V_1+O(|\psi|^3),
\end{equation}
\begin{equation}
  W=|\psi|^2W_2+O(|\psi|^4),
\end{equation}
\begin{equation}
  H=1+|\psi|^2 H_2+O(|\psi|^4).
\end{equation}
At leading order we find
\begin{equation}
  \rmd_\tau U_1-2V_1=\cos\tau,
\end{equation}
\begin{equation}
  \rmd_\tau V_1+(2-q)U_1=0,
\end{equation}
with the solution (having the required period of $2\pi$)
\begin{equation}
  U_1=\f{\sin\tau}{2q-3},
\end{equation}
\begin{equation}
  V_1=\f{(2-q)\cos\tau}{2q-3},
\end{equation}
provided that $q\ne{\textstyle\f{3}{2}}$.  This solution breaks down
in the Keplerian case because of the coincidence between the orbital
frequency and the epicyclic frequency.

Solutions valid for small $|\psi|$ and~$q$ close to
${\textstyle\f{3}{2}}$ can be found instead by the following
expansion:
\begin{equation}
  U=|\psi|^{-1}[U_0+|\psi|^2U_2+O(|\psi|^2)],
\end{equation}
\begin{equation}
  V=|\psi|^{-1}[V_0+|\psi|^2V_2+O(|\psi|^2)],
\end{equation}
\begin{equation}
  W=W_0+O(|\psi|^2),
\end{equation}
\begin{equation}
  H=H_0+O(|\psi|^2),
\end{equation}
\begin{equation}
  q=\f{3}{2}+|\psi|^2Q.
\end{equation}
Then we find
\begin{equation}
  \rmd_\tau U_0-2V_0=0,
\label{duu0}
\end{equation}
\begin{equation}
  \rmd_\tau V_0+\f{1}{2}U_0=0,
\label{dvv0}
\end{equation}
\begin{equation}
  \rmd_\tau W_0+\sin\tau\,U_0=H_0^{-1}-H_0,
\label{dww0}
\end{equation}
\begin{equation}
  \rmd_\tau H_0=W_0+\cos\tau\,U_0,
\label{dhh0}
\end{equation}
\begin{equation}
  \rmd_\tau U_2-2V_2=\cos\tau\,H_0^{-1},
\label{duu2}
\end{equation}
\begin{equation}
  \rmd_\tau V_2+\f{1}{2}U_2-QU_0=0.
\label{dvv2}
\end{equation}

Equations~(\ref{duu0}) and~(\ref{dvv0}) give
$(\rmd_\tau^2+1)(U_0,V_0)=0$ (a free linear epicyclic oscillator) and
have the general solution
\begin{equation}
  V_0=A\cos\tau+B\sin\tau,
\end{equation}
\begin{equation}
  U_0=2A\sin\tau-2B\cos\tau.
\end{equation}
We set $B=0$ on grounds of symmetry.

Equations~(\ref{dww0}) and~(\ref{dhh0}) give
\begin{equation}
  \rmd_\tau^2H_0=H_0^{-1}-H_0+A(3\cos2\tau-1).
\label{h0}
\end{equation}
Here we see the nonlinear vertical oscillator forced by the coupling
(through the warp) to the horizontal oscillator.  The natural
frequency of the vertical oscillator in the linear regime is
$\sqrt{2}$, which is far from resonance with the forcing at
frequency~$2$.  However, the natural frequency depends on the
amplitude of the oscillation.  The constant term $-A$ in the forcing
also shifts the centre of the oscillation away from the equilibrium
position $H_0=1$.

Equations~(\ref{duu2}) and~(\ref{dvv2}) give
\begin{equation}
  (\rmd_\tau^2+1)V_2=2QA\cos\tau-\f{1}{2}\cos\tau\,H_0^{-1},
\label{v2}
\end{equation}
a forced linear epicyclic oscillator.  For a periodic solution to
exist, the right-hand side must contain no Fourier component
proportional to $\cos\tau$.  The mean value and the $\cos2\tau$
content of $H_0^{-1}$ depend nonlinearly on $A$.  The aforementioned
solvability condition is therefore a nonlinear equation for $A$, which
may or may not have a solution.

\begin{figure}
\centerline{\epsfysize8cm\epsfbox{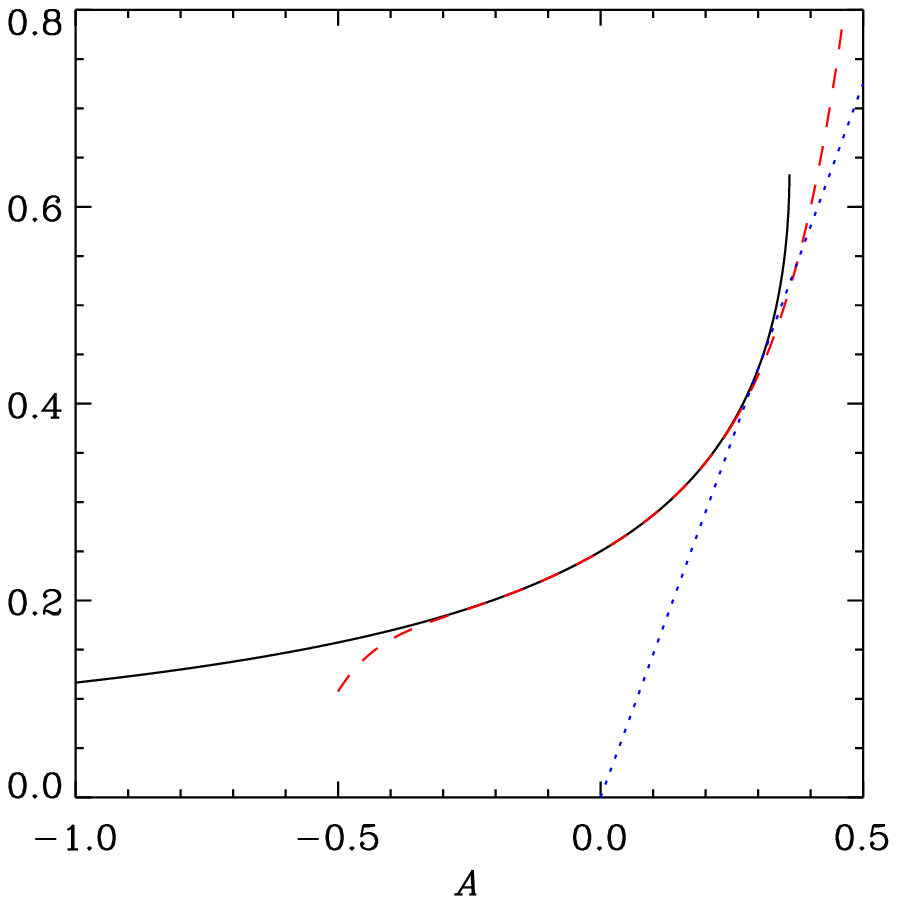}}
\caption{The numerically determined function of the amplitude $A$ that
  should equal $QA$ when the solvability condition for
  equation~(\ref{v2}) is applied.  The black solid line, which
  terminates at $A\approx0.36$, shows the component of $\cos\tau$ in
  the Fourier series of $(1/4)\cos\tau\,H_0^{-1}$ determined from the
  numerical solution of equation~(\ref{h0}).  The red dashed line
  shows the series approximation given by the right-hand side of
  equation~(\ref{qa_series}).  The blue dotted line is of slope $1.45$
  and passes through the origin.}
\label{f:solvability}
\end{figure}

For sufficiently small $A$, the solution of equation~(\ref{h0}) can be
obtained by a series expansion.  It is
\begin{eqnarray}
  \lefteqn{H_0=1-\f{A}{2}(1+3\cos2\tau)+\f{A^2}{112}(77-84\cos2\tau-9\cos4\tau)}&\nonumber\\
  &&\qquad+O(A^3),
\label{h0_series}
\end{eqnarray}
and so
\begin{eqnarray}
  \lefteqn{2QA\cos\tau-\f{1}{2}\cos\tau\,H_0^{-1}=-\f{1}{2}\cos\tau}&\nonumber\\
  &&+\f{A}{8}[(16Q-5)\cos\tau-3\cos3\tau]\nonumber\\
  &&-\f{A^2}{448}(406\cos\tau+387\cos3\tau+135\cos5\tau)\nonumber\\
  &&+O(A^3).
\end{eqnarray}
This series can be continued to higher order and is accurate if $|A|$
is not too large.  The solvability condition gives
\begin{eqnarray}
  \lefteqn{QA=\f{1}{4}+\f{5A}{16}+\f{29A^2}{64}+\f{333A^3}{448}+\f{270913A^4}{200704}}&\nonumber\\
  &&+\f{15848541A^5}{5970944}+\f{36001767715A^6}{6496387072}\nonumber\\
  &&+\f{253686124482669A^7}{20969525420032}+O(A^8).
\label{qa_series}
\end{eqnarray}
For larger $|A|$, we solve equation~(\ref{h0}) numerically instead.
The solution breaks down for $A\ga0.36$, although it can be continued
to large negative $A$.  (In fact the solution of equation~\ref{h0} is
not unique, but we consider here only the solution that is consistent
with the expansion~\ref{h0_series}.)  In Fig.~\ref{f:solvability} we
plot the numerically determined function of~$A$ that should equal $QA$
when the solvability condition is applied.  The red dashed line shows
the series approximation given by the right-hand side of
equation~(\ref{qa_series}).  The solvability condition is satisfied
when the black line in Fig.~\ref{f:solvability} intersects with a
straight line of slope $Q$ through the origin.  For negative values
of~$Q$ this condition can always be met for some negative value
of~$A$, but for positive values of~$Q$ a solution exists only if $Q$
is sufficiently large.  The critical slope is indicated by the blue
dotted line and corresponds to $Q\approx1.45$.  Therefore there are no
solutions for $0<Q\la1.45$.  In the full numerical problem for
inviscid laminar flows, the solution does indeed break down at
$Q\approx1.45$ and $A\approx0.30$.

The breakdown of the solution can be understood as a type of nonlinear
resonance.  It occurs while the nonlinear vertical oscillator exhibits
a finite response, but the anharmonic and asymmetric character of that
oscillator is essential to the behaviour.

\label{lastpage}


\begin{thebibliography}{}

\bibitem[\protect\citeauthoryear{Bardeen 
\& Petterson}{1975}]{1975ApJ...195L..65B} Bardeen J.~M., Petterson J.~A., 1975, ApJ, 195, L65 

\bibitem[\protect\citeauthoryear{Borderies, Goldreich 
\& Tremaine}{1989}]{1989Icar...80..344B} Borderies N., Goldreich P., Tremaine S., 1989, Icarus, 80, 344

\bibitem[\protect\citeauthoryear{Foucart 
\& Lai}{2011}]{2011MNRAS.412.2799F} Foucart F., Lai D., 2011, MNRAS, 412, 2799 

\bibitem[\protect\citeauthoryear{Foulkes, Haswell 
\& Murray}{2006}]{2006MNRAS.366.1399F} Foulkes S.~B., Haswell C.~A., Murray J.~R., 2006, MNRAS, 366, 1399 

\bibitem[\protect\citeauthoryear{Foulkes, Haswell 
\& Murray}{2010}]{2010MNRAS.401.1275F} Foulkes S.~B., Haswell C.~A., Murray J.~R., 2010, MNRAS, 401, 1275 

\bibitem[\protect\citeauthoryear{Fragile 
\& Anninos}{2005}]{2005ApJ...623..347F} Fragile P.~C., Anninos P., 2005, ApJ, 623, 347 

\bibitem[\protect\citeauthoryear{Fragile et 
al.}{2007}]{2007ApJ...668..417F} Fragile P.~C., Blaes O.~M., Anninos P., 
Salmonson J.~D., 2007, ApJ, 668, 417 

\bibitem[\protect\citeauthoryear{Fragile et 
al.}{2009}]{2009ApJ...691..482F} Fragile P.~C., Lindner C.~C., Anninos P., 
Salmonson J.~D., 2009, ApJ, 691, 482 

\bibitem[\protect\citeauthoryear{Fragner 
\& Nelson}{2010}]{2010A&A...511A..77F} Fragner M.~M., Nelson R.~P., 2010, A\&A, 511, A77 

\bibitem[\protect\citeauthoryear{Goldreich 
\& Lynden-Bell}{1965}]{1965MNRAS.130..125G} Goldreich P., Lynden-Bell D., 1965, MNRAS, 130, 125

\bibitem[\protect\citeauthoryear{Greenhill}{2005}]{2005ASPC..340..203G} 
Greenhill L.~J., 2005, in Romney, J.~D., Reid, M.~J., eds, Future Directions in High Resolution Astronomy, ASP Conf.\ Ser., 340, 203 

\bibitem[\protect\citeauthoryear{Hatchett, Begelman, 
\& Sarazin}{1981}]{1981ApJ...247..677H} Hatchett S.~P., Begelman M.~C., Sarazin C.~L., 1981, ApJ, 247, 677

\bibitem[\protect\citeauthoryear{Hawley, Gammie 
\& Balbus}{1995}]{1995ApJ...440..742H} Hawley J.~F., Gammie C.~F., Balbus S.~A., 1995, ApJ, 440, 742 

\bibitem[\protect\citeauthoryear{Katz}{1973}]{1973NPhS..246...87K} Katz 
J.~I., 1973, Nature, 246, 87

\bibitem[\protect\citeauthoryear{Kotze 
\& Charles}{2012}]{2012MNRAS.420.1575K} Kotze M.~M., Charles P.~A., 2012, MNRAS, 420, 1575 

\bibitem[\protect\citeauthoryear{Kumar 
\& Pringle}{1985}]{1985MNRAS.213..435K} Kumar S., Pringle J.~E., 1985, MNRAS, 213, 435

\bibitem[\protect\citeauthoryear{Lai}{1999}]{1999ApJ...524.1030L} Lai D., 
1999, ApJ, 524, 1030

\bibitem[\protect\citeauthoryear{Larwood}{1997}]{1997MNRAS.290..490L} 
Larwood J.~D., 1997, MNRAS, 290, 490

\bibitem[\protect\citeauthoryear{Larwood et 
al.}{1996}]{1996MNRAS.282..597L} Larwood J.~D., Nelson R.~P., Papaloizou 
J.~C.~B., Terquem C., 1996, MNRAS, 282, 597 

\bibitem[\protect\citeauthoryear{Larwood 
\& Papaloizou}{1997}]{1997MNRAS.285..288L} Larwood J.~D., Papaloizou J.~C.~B., 1997, MNRAS, 285, 288 

\bibitem[\protect\citeauthoryear{Lodato 
\& Price}{2010}]{2010MNRAS.405.1212L} Lodato G., Price D.~J., 2010, MNRAS, 405, 1212 

\bibitem[\protect\citeauthoryear{Lodato 
\& Pringle}{2007}]{2007MNRAS.381.1287L} Lodato G., Pringle J.~E., 2007, MNRAS, 381, 1287 

\bibitem[\protect\citeauthoryear{Lubow}{1992}]{1992ApJ...398..525L} Lubow 
S.~H., 1992, ApJ, 398, 525

\bibitem[\protect\citeauthoryear{Miyoshi et 
al.}{1995}]{1995Natur.373..127M} Miyoshi M., Moran J., Herrnstein J., 
Greenhill L., Nakai N., Diamond P., Inoue M., 1995, Nature, 373, 127 

\bibitem[\protect\citeauthoryear{Murray et al.}{2002}]{2002MNRAS.335..247M} 
Murray J.~R., Chakrabarty D., Wynn G.~A., Kramer L., 2002, MNRAS, 335, 247 

\bibitem[\protect\citeauthoryear{Nelson 
\& Papaloizou}{1999}]{1999MNRAS.309..929N} Nelson R.~P., Papaloizou J.~C.~B., 1999, MNRAS, 309, 929 

\bibitem[\protect\citeauthoryear{Nelson 
\& Papaloizou}{2000}]{2000MNRAS.315..570N} Nelson R.~P., Papaloizou J.~C.~B., 2000, MNRAS, 315, 570 

\bibitem[\protect\citeauthoryear{Nixon 
\& King}{2012}]{2012MNRAS.421.1201N} Nixon C.~J., King A.~R., 2012, MNRAS, 421, 1201 

\bibitem[\protect\citeauthoryear{Nixon, King
\& Price}{2012}]{2012MNRAS.422.2547N} Nixon C.~J., King A.~R., Price D.~J., 2012, MNRAS, 422, 2547 

\bibitem[\protect\citeauthoryear{Nixon et al.}{2012}]{2012ApJ...757L..24N} 
Nixon C., King A., Price D., Frank J., 2012, ApJ, 757, L24 

\bibitem[\protect\citeauthoryear{Ogilvie}{1999}]{1999MNRAS.304..557O} 
Ogilvie G.~I., 1999, MNRAS, 304, 557 

\bibitem[\protect\citeauthoryear{Ogilvie}{2000}]{2000MNRAS.317..607O} 
Ogilvie G.~I., 2000, MNRAS, 317, 607

\bibitem[\protect\citeauthoryear{Ogilvie}{2006}]{2006MNRAS.365..977O} 
Ogilvie G.~I., 2006, MNRAS, 365, 977 

\bibitem[\protect\citeauthoryear{Ogilvie
\& Latter}{2013}]{OL13} 
Ogilvie G.~I., Latter H.~N., 2013, submitted to MNRAS (Paper~II)

\bibitem[\protect\citeauthoryear{Papaloizou 
\& Lin}{1995}]{1995ApJ...438..841P} Papaloizou J.~C.~B., Lin D.~N.~C., 1995, ApJ, 438, 841

\bibitem[\protect\citeauthoryear{Papaloizou 
\& Pringle}{1983}]{1983MNRAS.202.1181P} Papaloizou J.~C.~B., Pringle J.~E., 1983, MNRAS, 202, 1181 

\bibitem[\protect\citeauthoryear{Papaloizou 
\& Terquem}{1995}]{1995MNRAS.274..987P} Papaloizou J.~C.~B., Terquem C., 1995, MNRAS, 274, 987 

\bibitem[\protect\citeauthoryear{Petterson}{1977}]{1977ApJ...214..550P} 
Petterson J.~A., 1977, ApJ, 214, 550 

\bibitem[\protect\citeauthoryear{Petterson}{1978}]{1978ApJ...226..253P} 
Petterson J.~A., 1978, ApJ, 226, 253 

\bibitem[\protect\citeauthoryear{Pringle}{1992}]{1992MNRAS.258..811P} 
Pringle J.~E., 1992, MNRAS, 258, 811 

\bibitem[\protect\citeauthoryear{Pringle}{1996}]{1996MNRAS.281..357P} 
Pringle J.~E., 1996, MNRAS, 281, 357

\bibitem[\protect\citeauthoryear{Schandl 
\& Meyer}{1994}]{1994A&A...289..149S} Schandl S., Meyer F., 1994, A\&A, 289, 149

\bibitem[\protect\citeauthoryear{Xiang-Gruess 
\& Papaloizou}{2013}]{2013MNRAS.431.1320X} Xiang-Gruess M., Papaloizou J.~C.~B., 2013, MNRAS, 431, 1320 

\end{thebibliography}
\end{document}